\documentclass[12pt,preprint]{aastex}
\usepackage{psfig,graphicx}
 
\newcommand{\apg}{\gtrsim}
\newcommand{\apl}{\lesssim}
\newcommand{\cmjj}{\mbox{${\rm cm^{-2}}$}}
\newcommand{\etal}{et al.}
\newcommand{\hI}{\mbox{H\,I}}

\newcommand{\kms}{\mbox{km\ s${^{-1}}$}}
\newcommand{\lya}{\mbox{${\rm Ly}\alpha$}}

\begin{document}
 
\lefthead{Chen \etal}
\righthead{ORIGIN OF DAMPED \lya\ ABSORBERS}

\slugcomment{Accepted by The Astrophysical Journal}

\title{ABUNDANCE PROFILES AND KINEMATICS OF DAMPED \lya\ ABSORBING GALAXIES AT 
$z<0.65$\altaffilmark{1,2,3}}

\author{Hsiao-Wen Chen\altaffilmark{4}}
\affil{Center for Space Research, Massachusetts Institute of 
Technology, Cambridge, MA 02139-4307, U.S.A. \\
hchen@space.mit.edu}

\and

\author{Robert C. Kennicutt, Jr.}
\affil{Steward Observatory, University of Arizona, Tucson, AZ 85721, U.S.A.\\
rkennicutt@as.arizona.edu}

\and

\author{Michael Rauch}
\affil{Carnegie Observatories, 813 Santa Barbara St, Pasadena,
CA 91101, U.S.A. \\
mr@ociw.edu}

\altaffiltext{1}{Observations reported here were obtained in part at the MMT 
Observatory, a joint facility of the University of Arizona and the Smithsonian 
Institution.}
\altaffiltext{2}{Observations reported here were obtained in part at the 
Magellan telescopes, a collaboration between the Observatories of the Carnegie 
Institution of Washington, University of Arizona, Harvard University, 
University of Michigan, and Massachusetts Institute of Technology.}
\altaffiltext{3}{Based in part on observations with the NASA/ESA Hubble Space
Telescope, obtained at the Space Telescope Science Institute, which is operated
by the Association of Universities for Research in Astronomy, Inc., under NASA
contract NAS5--26555.}
\altaffiltext{4}{Hubble Fellow}

\clearpage

\begin{abstract}

  We present a spectroscopic study of six damped \lya\ absorption (DLA) systems
at $z<0.65$, based on moderate-to-high resolution spectra of the galaxies 
responsible for the absorbers.  Combining known metallicity measurements of the
absorbers with known optical properties of the absorbing galaxies, we confirm 
that the low metal content of the DLA population can arise naturally as a 
combination of gas cross-section selection and metallicity gradients commonly 
observed in local disk galaxies.  We also study the Tully-Fisher relation of 
the DLA-selected galaxies and find little detectable evidence for evolution in 
the disk population between $z=0$ and $z\sim 0.5$.  Additional results of our 
analysis are as follows. (1) The DLA galaxies exhibit a range of spectral 
properties, from post-starburst, to normal disks, and to starburst systems, 
supporting the idea that DLA galaxies are drawn from the typical field 
population. (2) Large rotating \hI\ disks of radius $30\ h^{-1}$ kpc and of 
dynamic mass $M_{\rm dyn} > 10^{11}\,h^{-1}\,{\rm M}_\odot$ appear to be common
at intermediate redshifts. (3) Using an ensemble of six galaxy-DLA pairs, we 
derive an abundance profile that is characterized by a radial gradient of 
$-0.041\pm 0.012$ dex per kiloparsec (or equivalently a scale length of 
$10.6\ h^{-1}$ kpc) from galactic center to $30\ h^{-1}$ kpc radius.  (4) 
Adopting known $N(\hI)$ profiles of nearby galaxies and the best-fit radial
gradient, we further derive an $N(\hI)$-weighted mean metallicity $\langle Z
\rangle_{\rm weighted} = -0.50\pm 0.07$ for the DLA population over 100 random
lines of sight, consistent with $\langle Z\rangle_{\rm weighted} = 
-0.64_{-0.86}^{+0.40}$ observed for $z\sim 1$ DLA systems from Prochaska et 
al.  Our analysis demonstrates that the low metal content of DLA systems does 
not rule out the possibility that the DLA population trace the field galaxy 
population.

\end{abstract}


\newpage

\section{INTRODUCTION}

  Damped \lya\ absorption (DLA) systems are thought to probe the high-redshift
analogy of neutral gas regions that resembles the disks of nearby luminous 
galaxies and represent a unique sample for studying the interstellar medium 
(ISM) of distant galaxies at redshifts as high as the background QSOs can be 
observed.  These absorbers are typically selected to have neutral hydrogen 
column density $N(\hI) \ge 2\times 10^{20}$ \cmjj.  To date, substantial 
progress has been made to establish the chemical enrichment history in the cold
gas traced by the DLA systems.  Results obtained by various authors (e.g.\ 
Pettini \etal\ 1999; Prochaska \etal\ 2003a) indicate that the mean 
metallicities observed in the DLA systems are substantially lower than the mean
metal content observed in the solar neighborhood (e.g.\ Sofia \& Meyer 2001) at
all redshifts that have been studied.  

  The low metal content observed in DLA systems suggests that these absorbers
do not trace the bulk of star formation, particularly because galaxies at 
redshift $z<1$ exhibit a strong luminosity-metallicity correlation---more 
luminous galaxies are also more metal enriched (e.g.\ Kobulnicky \& Zaritsky 
1999; Tremonti \etal\ 2004).  Consequently, this would suggest that the DLA 
systems are not likely to be the progenitors of present-day luminous disk 
galaxies.  Recent analysis presented by Wolfe \etal\ (2003a,b) suggests, 
however, that the DLA systems may contribute equally to the star formation rate
density at $z=3$ as the luminous starburst population selected at rest-frame UV
wavelengths (e.g.\ Steidel \etal\ 1999).  Wolfe \etal\ further propose that the
discrepancy between high star formation rate and low metallicity in the DLA 
systems may be understood if star formation occurs in a compact region that has
little cross section.

  Identifying the stellar content of the DLA galaxies is clearly essential for 
understanding the nature of the absorbing galaxy population.  But this has
been a challenging task, because these galaxies are faint and located at small 
projected distances to the bright background QSOs (as implied by the intrinsic
high column density of the absorbers).  Despite extensive searches in the past 
decade, the number of known DLA galaxies has remained relatively small.  Of all
the 23 DLA systems found at $z\apl 1$, 13 have been identified with their host
galaxies using either spectroscopic or photometric redshift techniques (Rao \& 
Turnshek 1998; Lane \etal\ 1998; Miller, Knezek, \& Bregman 1999; Turnshek 
\etal\ 2001; Cohen 2001; Bowen, Tripp, \& Jenkins 2001; Chen \& Lanzetta 2003; 
Lacy \etal\ 2003), six systems have host candidates found in deep images 
(Steidel \etal\ 1994; Le Brun \etal\ 1997; Rao \etal\ 2003), and two systems
still remain unidentified after substantial search efforts (Steidel \etal\ 
1997; Cohen 2001).  Based on a study of eight DLA systems along seven lines of 
sight, Le Brun \etal\ (1997) have pointed out that galaxies giving rise to 
low-redshift DLAs display a wide range of morphologies.

  Chen \& Lanzetta (2003) studied the 11 galaxies that have been confirmed at 
the time to produce DLA systems and showed that typical $L_*$ galaxies at $z 
\sim 0.5$ are surrounded by an extended neutral gaseous envelope of radius 
$R_* = 24 - 30\ h^{-1}$ kpc at $N(\hI)=10^{20}$ \cmjj.  In addition, when a 
known galaxy luminosity function is adopted, the neutral gas cross-section 
weighted luminosity distribution of field galaxies would produce the observed 
luminosity distribution of DLA galaxies.  This result lends strong support for 
the idea that galaxies selected by association with known DLA systems are 
representative of the field galaxy population and that a large contribution of 
dwarfs ($M_B - 5\log h\ge -17$) to the total neutral gas cross section is not 
necessary.  But it also underscores the issue of the apparent low metallicity 
of the DLA population.

  We have been pursuing a spectroscopic study of $z<1$ galaxies known to 
produce DLA systems, in order to understand the low metallicity nature 
observed in the absorbers.  Here we present moderate resolution spectra of four
previously known DLA galaxies toward PKS0439$-$433 ($z_{\rm DLA} = 0.101$), 
Q0738$+$313 ($z_{\rm DLA}=0.221$), AO\,0235$+$164 ($z_{\rm DLA} = 0.524$), and 
B2\,0827$+$243 ($z_{\rm DLA}=0.525$).  We also add two DLA galaxies toward 
Q0809$+$483 ($z_{\rm DLA}=0.437$) and LBQS0058$+$0155 ($z_{\rm DLA}=0.613$) 
that are confirmed in our spectroscopic program.  Furthermore, metallicity 
measurements of the neutral gas are available from the literature for the DLA 
systems toward B2\,0827$+$243 (Khare \etal\ 2004) and LBQS0058$+$0155 (Pettini 
\etal\ 2000).  We present two new measurements for the DLA systems toward
PKS0439$-$433 and AO\,0235$+$164 based on the ultraviolet spectra of the 
background QSOs retrieved from the Hubble Space Telescope (HST) data archive.

  Combining known metallicity measurements of the absorbers with known optical
properties of the absorbing galaxies yields a galaxy-DLA pair sample that has 
two unique applications.  First, it allows us to go beyond simple cross-section
measurements and to establish an empirical correlation between the kinematics 
and metallicity of the absorbers and those of the absorbing galaxies.  A 
particularly interesting aspect of this application is to investigate whether 
or not the low metal content of the DLA population arises naturally as a 
combination of gas cross-section selection and metallicity gradients commonly 
observed in local disk galaxies (e.g.\ Zaritsky \etal\ 1994; Ferguson, 
Gallagher, \& Wyse 1998; van Zee \etal\ 1998).  Second, the sample offers a 
means of studying the disk population at intermediate redshift using galaxies 
selected uniformly based on known neutral gas content ($N(\hI)\apg 10^{20}$ 
\cmjj), rather than optical brightness or color.  Specifically, we will study
disk evolution based on a comparison of the Tully-Fisher relation of the DLA 
galaxies and those of field galaxies.

  We describe our observations and data reduction in \S\ 2 and present spectral
properties of the galaxy and absorber pairs toward individual sightlines in 
\S\ 3.  In \S\ 4, we investigate the Tully-Fisher relation of DLA selected 
galaxies and chemical abundance profiles of galactic disks.  Finally, a summary
is presented in \S\ 5.  Throughout the paper, we adopt a $\Lambda$ cosmology, 
$\Omega_{\rm M}=0.3$ and $\Omega_\Lambda = 0.7$, with a dimensionless Hubble 
constant $h = H_0/(100 \ {\rm km} \ {\rm s}^{-1}\ {\rm Mpc}^{-1})$.

\section{OBSERVATIONS AND DATA REDUCTION}

  We have obtained moderate-resolution spectra of six absorbing galaxies 
toward the lines of sight of PKS0439$-$433, Q0738$+$313, Q0809$+$483, 
AO\,0235$+$164, B2\,0827$+$243, and LBQS0058$+$0155, using the Magellan 
telescopes at Las Campanas and MMT on Mt.\ Hopkins.  These galaxies are known 
to give rise to strong \lya\ absorbers of $N(\hI)\apg 10^{20} \cmjj$ (e.g.\ 
Rao \& Turnshek 2000; Chen \& Lanzetta 2003).  The properties of the absorbing 
galaxies are summarized in Table 1.  High spatial resolution images of the 
galaxies toward PKS0439$-$433, Q0809$+$483, AO\,0235$+$164, B2\,0827$+$243, and
LBQS0058$+$0155 are presented in Figure 1, and these illustrate the relative 
locations of the galaxies with respect to the background QSOs and show their
visual morphology.  The image surrounding the field of PKS0439$-$433 was 
obtained using the Tek\#5 CCD imager with an $R$-band filter on the du Pont 
telescope at Las Campanas (Chen \& Lanzetta 2003).  The images surrounding 
Q0809$+$483, AO\,0235$+$164, B2\,0827$+$243, and LBQS0058$+$0155 were obtained 
using HST, the Wide Field and Planetary Camera 2 (WFPC2) with the F702W filter.
Raw images were downloaded from the HST data archive, and processed and coadded
using our own image processing programs (Chen \& Lanzetta 2003).  No image 
surrounding Q0738$+$313 is available in public data archives, but a deep 
$K$-band image is published in Turnshek \etal\ (2001).  These high spatial 
resolution images allowed us to measure accurately the inclination angles of 
disk galaxies (PKS0439$-$433, Q0809$+$483, and B2\,0827$+$243) for rotation 
curve studies.

  We have also searched through the HST data archive and located
moderate-resolution, ultraviolet and optical spectra of the background QSOs, 
PKS0439$-$433 and AO\,0235$+$164.  The spectra were obtained using either the 
Faint Object Spectrograph (FOS) or the Space Telescope Imaging Spectrograph 
(STIS).  They are suitable for studying the metal abundances of the neutral gas
using absorption features produced by heavy ions such as iron and/or manganese.
The observation and data reduction procedures for the spectra are described in 
the following sections.  A journal of the ground-based, long-slit spectroscopic
observations is listed in Table 2.  A journal of spectroscopic observations 
using HST is listed in Table 3.

\subsection{Magellan Observations}

  Long-slit spectra of the absorbing galaxies toward PKS0439$-$433 and 
LBQS0058$+$0155 were obtained using the Boller \& Chivens Spectrograph (B\&C) 
on the Clay telescope in September 2002.  The camera has a pixel scale of 0.25 
arcsec at the Nasmyth focus.  We used a 1.2 arcsec slit with a 600 l/mm grating
for a dispersion of 1.6 \AA\ per pixel, and the same slit width with a 1200 
l/mm grating for a dispersion of 0.8 \AA\ per pixel.  An order-blocking filter 
was inserted to exclude photons at wavelength $\lambda < 4000 \AA$.  For each 
slit setup in low-resolution mode, we aligned the slit along the parallactic 
angle to reduce differential slit loss due to atmospheric refraction.  In 
high-resolution mode, we aligned the slit along the major axis of disk galaxies
to obtain spatially resolved emission features.  The observations were carried 
out in a series of two to three exposures of between 600 and 1200 s each, and 
no dither was applied between individual exposures.  Calibration frames for 
flat-field correction and wavelength solutions were taken immediately before or
after each set of science exposures using a flat screen in front of the 
secondary mirror.  For the absorbing galaxy toward PKS0439$-$433, we set the 
grating at two tilt angles to cover a wavelength range from 3900 \AA\ through 
7500 \AA\ in low-resolution mode.  A spectrophotometric standard star was 
observed at the beginning and the end of each night for relative flux 
calibration.

\subsection{MMT Observations}

  Long-slit spectra of the absorbing galaxies toward Q0738$+$313, Q0809$+$483, 
AO\,0235$+$164 and B2\,0827$+$243 were obtained using the Blue Channel in the
MMT Spectrograph on the MMT in December 2003.  The camera has a pixel scale of 
0.3 arcsec at the Cassegrain focus of the telescope.  We used a 1 arcsec slit 
with a 500 l/mm grating to obtain a dispersion of 1.2 \AA\ per pixel and with a
1200 l/mm grating to obtain a dispersion of 0.5 \AA\ per pixel.  For each slit 
setup, we align the slit along the parallactic angle in low-resolution mode and
along the major axis of disk galaxies in high-resolution mode.  The 
observations were carried out in a series of three to four exposures of between
1200 and 1800 s each, and no dither was applied between individual exposures.
Calibration frames for flat-field correction and wavelength solutions were 
taken immediately after each set of science exposures using an internal quartz
lamp.  For the absorbing galaxy toward Q0738$+$313, we set the grating at two 
tilt angles to cover a wavelength range from 3900 \AA\ through 8200 \AA.  For 
the Q0809$+$483 sightline, the background QSO light was blended in the slit.  
We placed the slit on the QSO and obtained a 300-s exposure.  This frame was 
used to subtract the contaminating QSO light from the galaxy spectrum in the 
subsequent analysis.  A spectrophotometric standard was observed at the 
beginning and the end of each night for flux calibration.

\subsection{Reduction of Galaxy Spectra}

  All the spectroscopic data were processed following the standard data
reduction procedures.  Individual CCD frames were first bias subtracted using
available IRAF tasks and flattened using a median image of three spectroscopic
flat frames obtained for each pointing.  For low-resolution spectra, a single
spectrum was extracted and wavelength calibrated from each exposure.  By
using skylines in our science exposures the wavelength zero-point calibration 
was accurate to better than 0.5 \AA.  Wavelengths were further corrected to 
vacuum and heliocentric for comparisons with absorption-line measurements.
 
  Individual, extracted spectra were stacked to form a combined spectrum after
rejecting deviant pixels.  The stacked spectra were calibrated to an absolute 
flux scale using the sensitivity function derived from the spectrophotometric 
standards observed in each night.  No corrections for slit losses as a result
of a finite slit width was applied.  Consequently, the emission line fluxes 
may be underestimated using these spectra, but relative line flux ratios should
be accurate.  Comparisons of the continuum fluxes between blue and red grating 
tilts in the spectra obtained for PKS0439$-$433 and Q0738$+$313 showed that the
flux calibrations are consistent to within 5\%.  Finally, we corrected for the 
telluric A- and B-band absorption features at 7600 \AA\ and 6870 \AA\ in each 
stacked galaxy spectrum, using the much higher signal-to-noise ratio spectra of
the spectrophotometric standards.

  For the high-resolution spectra, individual frames were first registered 
using the continua and emission-line features and then stacked to form a 
combined two-dimensional spectrum.  We first applied a low-order polynomial fit
along the dispersion direction to remove the continua of the galaxy (and in the
case of Q0809$+$483 the background QSO).  Extended emission features due to 
H$\beta$ for PKS0439$-$433 and [O\,II] for Q0809$-$483 and B2\,0827$+$243 
became prominent.  We extracted one-dimensional spectra around the targeted 
emission features from individual rows, and estimated the line centroid per 
spectrum by fitting a Gaussian profile to the data to determine a velocity 
offset of the emission-line region along the slit.  Adjacent spectra were 
correlated between one another due to atmospheric seeing, and therefore the 
observed inner slope of the rotation curves of the galaxies is expected to be 
shallower than the intrinsic shape.  Nevertheless, the observations allow us to
compare for each system the relative kinematics between the ionized gas of the 
ISM and the neutral gas of the absorber along the disk and to determine the 
disk rotation speed.

\subsection{HST Observations}

  PKS0439$-$433 was first observed with HST using FOS and the G190H and G270H 
gratings, as part of the Quasar Absorption Line Key Project (Bahcall \etal\ 
1993).  A combined FOS spectrum of the QSO covering $1600-3277$ \AA\ (Evans \& 
Koratkar 2004) was retrieved from the HST archive for absorption line analysis.
The combined spectrum has a dispersion of $\sim 0.4$ \AA\ per pixel and a mean 
signal-to-noise ratio (SNR) of $\approx 20$ per resolution element between 2500
\AA\ and 2900 \AA, where we expect to detect the Fe\,II transitions from the 
DLA at $z=0.101$.  A single spectrum of the QSO, covering $1120-1717$ \AA, was 
later obtained using STIS and the G140L grating for a total exposure time of 
2510 s.  We retrieved the extracted one-dimensional spectrum from the HST 
archive for an accurate measurement of $N(\hI)$.  The spectrum has a dispersion
of 1.6 \AA\ per pixel and a mean SNR of $\approx 20$ per resolution element 
between 1250 \AA\ and 1400 \AA, where we expect to detect the \lya\ transition 
of the absorber.

  AO\,0235$+$164 was observed with HST using STIS the G430L and G750L gratings,
each in single exposures.  Additional spectra were obtained using STIS and the 
G230L grating to cover a broad spectral range from ultraviolet through 
optical wavelengths.  A series of five exposures were obtained in the G230L
setup.  Individual one-dimensional spectra were coadded to form an averaged 
spectrum and a 1 $\sigma$ error array using our own reduction program.  The
single spectrum obtained with the G430L grating has a dispersion of 2.6 \AA\ 
per pixel and a mean SNR of $\approx 30$ at 3600 \AA\ and $\approx 40$ at
3900 \AA, where we expect to identify a series of metal absorption lines such
as Fe\,II and Mn\,II.

\section{SPECTRAL PROPERTIES OF INDIVIDUAL DLAS}

  In this section we discuss each of the six DLA systems individually and
compare the properties of the neutral gas with the ISM properties of the 
absorbing galaxy, including kinematics and chemical abundances.  We have 
obtained Fe abundance measurements for the DLA systems toward PKS0439$-$433 
and AO\,0235$+$164, using available ultraviolet spectra from the HST archive.  
We have also measured a precise redshift of the absorbing galaxies using a 
$\chi^2$ analysis that compares the observed galaxy spectrum with models 
established from a linear combination of four principle-component spectra 
derived from the Sloan Digital Sky Survey galaxy 
sample\footnote{See http://spectro.princeton.edu/.}.  

\subsection{The DLA at $z=0.101$ Toward PKS0439$-$433}

  The strong metal-line absorber at $z=0.101$ toward PKS0439$-$433 has been
discussed in detail by Chen \& Lanzetta (2003).  The system was estimated to 
have $N(\hI)\sim 1\times10^{20}$ \cmjj\ based on an X-ray spectrum of the QSO 
by Wilkes \etal\ (1992).  A disk galaxy was found at the redshift of the DLA,
$5.1\ h^{-1}$ kpc projected distance away from the QSO line of sight.

\subsubsection{Properties of The Neutral Gas}

  The $N(\hI)$ measured from X-ray observations has a large uncertainty because
of uncertain foreground hydrogen column density along the line of sight in the 
Galaxy.  We have used the ultraviolet spectra of the QSO from HST FOS and STIS 
over 1120--3200 \AA\ to improve the accuracy of the $N(\hI)$ measurement, as 
well as to estimate the Fe\,II column density, $N({\rm Fe\,II})$.

  We first estimated $N({\rm Fe\,II})$ and the velocity width $b$ of Fe\,II
based on the absorption equivalent widths of five Fe\,II transitions published 
in Bechtold \etal\ (2002).  The results from a curve-of-growth analysis are 
presented in Figure 2.  We derived a best-fit $\log N({\rm Fe\,II})=14.58\pm 
0.06$ for a velocity width of $b=52\pm 3\,{\rm km} {\rm s}^{-1}$.  The errors 
represent the 95\% confidence interval.  The large best-fit $b$ suggests that 
there are multiple absorbing components in the gaseous cloud that are not 
resolved in the low-resolution FOS data.  The top five panels of Figure 3 shows
a comparison between the five observed Fe\,II absorption profiles and Voigt 
models produced using the best-fit $N({\rm Fe\,II})$, $b$, and the respective 
oscillator strengths.  Despite the low resolution of the QSO spectrum, our 
Fe\,II column density estimate based on a curve-of-growth analysis is correct 
(Jenkins 1986).

  Next we analyzed the STIS G140L spectrum of the QSO for estimating $N(\hI)$.
A close examination of the hydrogen \lya\ absorption profiles of the Galaxy and
of the DLA at $z=0.101$ shows non-zero fluxes in the cores of the lines, 
suggesting a background subtraction error in the spectrum extraction pipeline.
We applied a zeroth-order correction to the flux level in order to force zero 
fluxes in the core of the two damped absorbers, and derived a best-fit 
$\log\,N(\hI)=19.85\pm 0.10$ using the 
VPFIT\footnote{http://www.ast.cam.ac.uk/\char'176rfc/vpfit.html} software 
package.

  The revised \hI\ column density measurement places this absorber outside the
canonical $N(\hI)$ threshold of DLAs.  We performed a simple photoionization 
analysis using the CLOUDY software package (Ferland 2001) to evaluate the 
ionization state of the absorbing gas.  Adopting a nominal background ionizing 
photon flux at 912 \AA, $J_{912} \approx 6\times 10^{-23}$ ergs sec$^{-1}$ 
cm$^{-2}$ Hz$^{-1}$ Sr$^{-1}$ at $z<1$ (Scott \etal\ 2002), we found that the 
absorber may be characterized by an ionization fraction ${\rm H}^+/{\rm H} < 
0.1$ for a hydrogen number density $n_{\rm H}\ge 0.1$ cm$^{-3}$.  Our
photoionization analysis indicates that at $z=0.1$ the absorbing gas is likely 
to remain mostly neutral even with a lower $N(\hI)$, primarily because the 
metagalactic ionizing radiation field is nearly an order of magnitude weaker 
at $z<1$ than it was at $z\sim 3$ (Scott \etal\ 2000).  We therefore include 
this absorber in our DLA sample. 

  Finally, we combined the $N(\hI)$ and $N({\rm Fe\,II})$ measurements to 
derive $[{\rm Fe}/{\rm H}]$\footnote{$[{\rm Fe}/{\rm H}]\equiv\log ({\rm 
Fe}/{\rm H}) - \log ({\rm Fe}/{\rm H})_\odot$}$= -0.72\pm 0.12$ on the 
assumption that all iron is singly ionized and in neutral gaseous clouds 
(e.g.\ Vladilo \etal\ 2001; Prochaska \etal\ 2002).  The observed Fe abundance 
indicates $1/5$ solar metallicity in the DLA, but the actual metallicity of the
neutral gas is expected to be higher because Fe is susceptible to dust 
depletion.  Using a sample of 46 DLAs with available Zn and Fe abundance 
measurements (Pettini \etal\ 2000; Prochaska \etal\ 2003b), we obtained a 
median $\langle[{\rm Zn}/{\rm Fe}]\rangle=0.58\pm 0.28$, where the error 
represents the 1-$\sigma$ dispersion of the sample.  Because Zn is essentially 
undepleted in the ISM of the Galaxy, we derived a metallicity $Z=-0.14 \pm 
0.30$ for the DLA at $z=0.101$ toward PKS0439$-$433 based on the $[{\rm 
Fe}/{\rm H}]$ measurement and the median $[{\rm Zn}/{\rm Fe}]$ ratio.

\subsubsection{ISM Properties of The Absorbing Galaxy}

  The summed, extracted spectrum of the absorbing galaxy presented in Figure 4 
displays multiple emission and absorption features that suggest a young stellar
population and metal-enriched ISM.  The spectrum covers 3550--6850 \AA\ in the 
rest frame of the galaxy and shows [O\,II], H$\beta$, H$\alpha$, [N\,II], and 
[S\,II] emission features as well as G-band, Ca\,II H\&K, and Balmer absorption
line series, typical of what is seen in an intermediate-type disk galaxy.  We
present line flux measurements and corresponding errors of the identified 
emission features in Table 4.  The listed H$\alpha$ and H$\beta$ line fluxes 
have been corrected for stellar absorption by comparing the observed spectrum 
with theoretical models produced by the Bruzual \& Charlot (2003) stellar 
population synthesis code.  The redshift of the galaxy is $z=0.10104\pm 
0.00007$, $\approx 44$ \kms\ away from the redshift of the DLA.

  We examined the nature of the emission line region in the ISM using various
spectral diagnostics.  First, we estimated the amount of dust extinction 
internal to the absorbing galaxy using the observed H$\alpha$ and H$\beta$ line
flux ratio.  We derived a color excess $E(B-V)=0.22\pm 0.02$, following the 
formula presented in Calzetti, Kinney, \& Storchi-Bergmann (1994).  The 
Galactic extinction toward the sightline is $E(B-V)=0.016$ (Schlegel, 
Finkbeiner, \& Davis 1998), which is negligible in comparison to our derived 
color excess.  We therefore attributed all of the observed reddening to the 
absorbing galaxy, and corrected the observed line flux measurements using a 
Galactic extinction law (Cardelli, Clayton, \& Mathis 1989).  The extinction 
corrected emission-line luminosity is listed in the last column of Table 4.

  In order to characterize the type of line emission, we measured the 
extinction corrected [O\,III]\,$\lambda$5007/H$\beta$ and 
[N\,II]\,$\lambda$6583/H$\alpha$ line ratios.  We found 
$\log$\,[O\,III]\,$\lambda$5007/H$\beta = -0.50 \pm 0.01$ and
$\log$\,[N\,II]\,$\lambda$6583/H$\alpha = -0.38 \pm 0.01$.  These flux ratios 
are consistent with those measured for local H\,II regions, indicating that 
star formation, not AGN activity, is the dominating mechanism for producing 
these spectral features (Baldwin, Phillips, \& Terlevich 1981; Kennicutt \& 
Garnett 1996).  We also measured an extinction corrected
[N\,II]\,$\lambda$6584/[O\,II]\,$\lambda\lambda$3726,3729 line ratio and the
$R_{23}=$\,([O\,II]\,$\lambda\lambda\,3726,3729$$+$[O\,III]\,$\lambda\lambda\,4959,5007$)/H$\beta$ index.  We found 
$\log$\,[N\,II]\,$\lambda$6584/[O\,II]\,$\lambda\lambda3726,3729=0.03\pm0.02$
and $\log\,R_{23}=0.20\pm 0.02$.  These numbers together indicate that these
emission features originate in stellar photoionized H\,II regions.  

  Finally, we determined the interstellar oxygen abundance of the absorbing 
galaxy using several emission-line calibrators.  We first adopted the 
semi-empirical correlation between $12 + \log({\rm O}/{\rm H})$ and 
$\log$\,[N\,II]/H$\alpha$ from Pettini \& Pagel (2004) and derived $12 + 
\log({\rm O}/{\rm H}) = 8.68 \pm 0.18$\footnote{In the same paper, Pettini \&
Pagel also drew attention to the index 
$O3N2\equiv\log\{({[\rm O\,III]}\,\lambda5007/{\rm H}\beta)/
({\rm [N\,II]}\,\lambda6583/{\rm H}\alpha)\}$ as an alternative for estimating
the oxygen abundance in high-metallicity regime (greater than solar).  Adopting
the $O3N2$ index, we derived $12 + \log({\rm O}/{\rm H}) = 8.77 \pm 0.14$ for
the ISM of the absorbing galaxy.}  The result clearly shows that the ISM is 
metal rich.  The observed [N\,II]/H$\alpha$ ratio places the galaxy on the 
upper branch of the $R_{23}$-metallicity relation.  Using the calibration of 
McGaugh (1991), parameterized by Kobulnicky, Kennicutt, \& Pizagno (1999), we 
found $12 + \log({\rm O}/{\rm H}) = 9.19 \pm 0.15$, where the error is 
dominated by the uncertainty of the calibrator as reported by these authors.  
In summary, we found that the ISM of the DLA galaxy has been enriched to at 
least solar abundance and perhaps three times higher\footnote{We have adopted 
the revised oxygen abundance for the Sun, $12 + \log({\rm O}/{\rm H}) = 8.66$ 
(Allende-Prieto, Lambert, \& Asplund 2001; Asplund \etal\ 2004).}.  

  Comparison of the metallicities measured in the ISM and the neutral gas at
$\rho=5.1\ h^{-1}$ kpc shows an abundance decrement along the disk of the
galaxy (see \S\ 4.2 for discussions on possible systematic uncertainties).  A 
large metallicity gradient is often observed in local disk galaxies (e.g.\ 
Zaritsky, Kennicutt, \& Huchra 1994; Kennicutt, Bresolin, Garnett 2003), but 
has not been confirmed in DLA galaxies (e.g.\ Schulte-Ladbeck \etal\ 2004).

\subsubsection{Kinematic Properties of The DLA}

  A rotation curve analysis of DLA galaxies allows us to examine the kinematics
of the ISM and compare the motion of the absorber relative to the absorbing 
galaxy.  To measure the rotation curve of the absorbing galaxy toward 
PKS0439$-$433 along the stellar disk, we first analyzed the surface brightness
profile observed in a $K$-band image from Chen \& Lanzetta (2003).  We found 
that the stellar disk can be characterized with an inclination angle $i=58\pm 
5$ degrees and a position angle of the major axis $\alpha = 53\pm 2$ degrees, 
measured north through east.  Next, we measured the observed velocity shear 
$v_{\rm obs}$ as a function of projected distance $\rho$ from the center of the
galaxy, using the spatially resolved H$\beta$ emission feature.  Finally, the 
observed rotation distribution was deprojected to the plane along the stellar 
disk by converting $\rho$ to galactocentric radius $R$ following
\begin{equation}
\frac{R}{\rho}=\sqrt{1+\sin^2(\phi-\alpha)\tan^2(i)},
\end{equation}
and by converting $v_{\rm obs}$ to rotation speed along the disk following
\begin{equation}
v=\frac{v_{\rm obs}}{\cos(\phi-\alpha)\sin(i)}\sqrt{1+\sin^2(\phi-\alpha)\tan^2(i)},
\end{equation}
where $\phi$ is the slit orientation angle measured north through east.  The 
long-slit spectroscopic observations were carried out with the slit aligned 
along the major axis, i.e. $\phi = \alpha$.  We therefore have $R = \rho$ and 
$v=v_{\rm obs}/\sin i$, when deprojecting the observed rotation velocities 
determined from emission-line measurements of the galaxy.  Figure 5 shows the 
inclination corrected rotation curve, $v_{\rm obs}/\sin i$, in solid points 
measured for the absorbing galaxy.

  The velocity offset between the DLA ($z_{\rm DLA} = 0.10088$) and the 
systematic velocity of the absorbing galaxy ($z_{\rm gal} = 0.10104$) places an
additional point at large galactocentric radius that allows us to examine the 
kinematic nature of the absorber.  Figure 1 shows that the background QSO 
intersects the extended stellar disk at $\rho = 5.3\,h^{-1}$ kpc and $\phi =
-11.3$ degrees, leading to $R=9.5\pm 1.7\,h^{-1}$ kpc and $v_{\rm DLA} =
-233\pm 55$ \kms\ for the absorbing cloud.  The errors include uncertainties in
$i$ and $\alpha$ of the stellar disk, as well as uncertainties in the 
wavelength calibration of the HST/FOS spectrum.  The derived rotation velocity
of the DLA is shown as the open circle in Figure 5.  Comparison of the rotation
motion of the stellar disk and the DLA shows that the absorbing cloud is moving
along with the optical disk and suggests that the galaxy possesses an extended 
neutral gaseous disk to at least $9\,h^{-1}$ kpc radius.

\subsection{The DLA at $z=0.2212$ Toward Q0738$+$313}

  The DLA system with $N(\hI) = (7.9 \pm 1.4)\times 10^{20}$ \cmjj\ at $z = 
0.2212$ toward Q0738$+$313 (OI363) was selected by the presence of a 
Mg\,II absorption doublet and confirmed with subsequent HST observations using 
FOS (Rao \& Turnshek 1998).  The absorbing cloud is also detected in 21\,cm 
observations (Lane \etal\ 1998).  No additional metal absorption features have
been found with the absorber due to a lack of high spectral resolution data.  
Rao \& Turnshek (1998) also reported the detection of a compact galaxy  
$13.5\ h^{-1}$ kpc away from the QSO line of sight with $M_{\rm AB}(B) - 
5\,\log\,h=-17.7$.

  Figure 6 presents the summed, extracted spectrum, covering 3200--5300 \AA\ in
the rest frame of the galaxy.  There are no obvious emission-line features but 
we do observe a pronounced flux discontinuity at 4000 \AA, as well as strong 
absorption features from the G-band, Ca\,II H\&K, and Balmer series 
transitions.  We measured a $3\,\sigma$ upper limit over a resolution element 
($\approx 3.6$ \AA) to the [O\,II] line flux at $2.2 \times 10^{-18}\,{\rm 
erg}/{\rm sec}/{\rm cm}^2$ (Table 5).  The lack of strong [O\,II] emission to 
this sensitive limit suggests that either the galaxy is fairly quiescent with 
little on-going star formation or it is heavily reddened by dust.  The 
relatively flat spectral slope at rest-frame ultraviolet wavelengths and the 
strong 4000-\AA\ discontinuity together argue against dust reddening.  The 
redshift of the galaxy is $z=0.2222\pm 0.0001$, $\approx 250$ \kms\ away from 
the redshift of the DLA.

  The fact that a pronounced 4000-\AA\ flux discontinuity and strong Balmer 
series absorption dominate the spectral features of the galaxy also suggests 
that the last episode of star formation occurred in the recent past.  We 
attempted to constrain the age of the youngest stellar population in the galaxy
based on the observed absorption-line strengths.  First, we generated a grid of
Bruzual \& Charlot (2003) model spectra at different ages, assuming a Salpeter 
initial mass function.  The models spanned a range of star formation history 
from an instantaneous single burst to an exponentially declining star formation
rate (SFR) history with an e-folding time of $\tau=0.3$ Gyr (typical dynamic
time scale of an $L_*$ galaxy), and a range of metallicity from $1/5$ solar to 
solar.

  Next, we performed a $\chi^2$ analysis, in which the probability that the 
observed galaxy spectrum is consistent with a model template $k$ at a given age
$t$ is defined as the following,
\begin{equation}
p(k,t)=\prod_{i=1}\frac{1}{\sqrt{2\pi}\sigma_i}\exp\left[-\frac{1}{2}\frac{(f_i-F_i(k))^2}{\sigma_i^2}\right],
\end{equation}
where $i$ is the number of spectral elements, $f_i$ and $\sigma_i$ are the
observed flux and flux error in element $i$, and $F_i(k)$ is the corresponding
predicted flux from model $k$.  Finally, we maximized $p(k,t)$ with respect to 
stellar synthesis model $k$ at each given age to form $p(t)$ and then maximized
$p(k)$ with respect to age to find the best-fit age $t_*$.  We found that the 
observed spectrum is best described by an exponentially declining SFR model 
with a solar metallicity and $t_*=2.6_{-0.8}^{+0.3}$ Gyr.  The errors represent
the 1-$\sigma$ confidence interval over the models considered, and the best-fit
model spectrum is presented in Figure 6 (magenta curve).  

  Two points are learned from the stellar population analysis.  First, our 
analysis shows that the stellar population of the galaxy is best-described with
a solar metallicity.  Limiting the metallicity to $< 1/5$ solar does not 
produce a better fit and would increase the best-fit age to $> 9$ 
Gyr\footnote{At $z=0.22$, the universe is 10.8 Gyr old for the adopted 
$\Lambda$-cosmology and $H_0=70$ km s$^{-1}$ Mpc$^{-1}$.} at $z = 0.2222$,
which suggests a formation redshift of $z_f>5$ and is therefore not likely to 
be true.  Including dust reddening is likely to reduce the metallicity, but we 
note that a luminosity weighted metallicity of approximately solar is also 
expected from field populations (e.g.\ Jerjen, Binggeli, \& Barazza 2004).  
Second, including errors in the best-fit age, our analysis shows that major 
star formation in the galaxy terminated at least 1.5 Gyr ago.  On the other 
hand, the presence of a strong DLA indicates that the galaxy still possesses 
extended neutral gas, i.e.\ star forming reservoir.  The lack of on-going 
star formation suggests that the DLA arises in a relatively warm neutral 
medium; this is consistent with the spin temperature ($T_s \sim 1200$ K) 
derived from 21\,cm observations for the absorber (Lane \etal\ 1998).

\subsection{The DLA at $z=0.437$ Toward Q0809$+$483}

  The DLA at $z=0.4368$ toward Q0809$+$483 was selected based on a 21\,cm 
absorption line observed against the background radio-loud QSO (Brown \&
Mitchell 1983) and confirmed to have $\log N(\hI)/\cmjj = 20.8\pm 0.2$ in 
subsequent HST spectroscopy using FOS (Boiss\'e \etal\ 1998).  No additional 
metal absorption features have been found due to the limited spectral 
resolution and sensitivity of the data.  A spiral galaxy $1.5''$ away from the 
QSO line of sight with $AB({\rm F702W})=19.9$ has been reported as a candidate 
host galaxy for the DLA by Le Brun \etal\ (1997).

\subsubsection{ISM Properties of The Absorbing Galaxy}

  We have extracted a spectrum of the candidate DLA galaxy, and a separate 
spectrum taken at the QSO.  Both spectra cover 3600--5200 \AA\ at $z=0.437$.  
As it turned out, not only is the galaxy spectrum contaminated by the QSO 
light, but the H$\beta$ feature expected at $z = 0.437$ of the galaxy coincides
with the [O\,II] emission feature of the background QSO at $z_{\rm em}=0.871$.
It became necessary to subtract the blended QSO light before proceeding with 
further analysis.  

  To estimate the amount of contaminating QSO light, we first compared the line
fluxes observed in various features originating in the QSO.  The bottom two 
panels of Figure 7 show the summed, extracted spectra of the absorbing galaxy 
and the background QSO.  The [Ne\,V] $\lambda\lambda 3346,3426$ doublet and 
[Ne\,III] $\lambda 3868$ emission line from the QSO are apparent in both 
panels, in addition to the broad Mg\,II $\lambda 2799$ emission.  We adopted 
the QSO spectrum as a template and rescaled it to fit the Ne line fluxes 
observed in the contaminated galaxy spectrum.  The difference spectrum is 
presented in the top panel of Figure 7.  It shows that both the broad-band QSO 
features and narrow Ne emission lines are no longer present, leaving an 
emission-line dominated spectrum typical of a star-forming galaxy.  The 
residual fluxes in the difference spectrum at the locations of all the neon 
transitions are consistent with zero line fluxes, lending strong confidence in 
the detection of the H$\beta$ emission line in the difference spectrum.  

  The redshift of the galaxy is $z=0.43745\pm 0.00010$, $\approx 138$ \kms\ 
away from the redshift of the DLA.  We have confirmed the identification of the
galaxy as the DLA host.  At this redshift, the galaxy is $\rho=5.9\,h^{-1}$ kpc
from the QSO line of sight and has $M_{\rm AB}(B) - 5\,\log\,h=-20.3$.

  We measured line fluxes for [O\,II] and H$\beta$, and placed a $3\,\sigma$ 
upper limit over a resolution element ($\approx 3.6$ \AA) on the [O\,III] 
$\lambda\lambda 4959,5007$ line fluxes.  The results are listed in Table 6.  We
were unable to assess the dust content using the H$\alpha$ and H$\beta$ line 
ratio.  In the absence of dust obscuration, we derived an upper limit to the 
$R_{23}$ index for metallicity estimate and found $\log\,R_{23} \le -0.12$.  
This constrains the oxygen abundance in the ISM of the galaxy to be $12 + 
\log({\rm O}/{\rm H}) \ge 9.1$ for the upper branch in the metallicity-$R_{23}$
correlation or $12 + \log({\rm O}/{\rm H}) \le 7.0$ for the lower branch 
(Kobulnicky, Kennicutt, \& Pizagno 1999).  Taking into account that the 
[O\,III] $\lambda\lambda 4959, 5007$ lines are weak (not detected), we consider
the low metallicity solution unlikely and conclude that the ISM of the galaxy 
has an oxygen abundance of $12 + \log({\rm O}/{\rm H}) \approx 9.1$ or twice 
solar.  The high metallicity of the ISM is also supported by the 
luminosity-metallicity relation observed in nearby spiral galaxies (Kobulnicky 
\& Zaritsky 1999).

  Finally, we note that narrow Ca\,II H\&K absorption doublet at $z=0.437$ is 
observed in the spectrum of the background QSO at $-21\pm 15$ \kms\ from the 
systematic redshift of the galaxy.  The Ca\,II absorber appears to be moving
in the same direction with the DLA along the optical disk (see discussion in 
the next section) but has a substantial velocity lag along the line of sight 
(in comparison to $-138\pm 40$ \kms\ of the DLA with respect to the galaxy).  
Cold gas traced by Ca\,II is commonly observed in the local ISM and in the 
Galactic halo (e.g.\ Welty, Morton, \& Hobbs 1996), and exhibits complex 
velocity structures.  A large fraction of the Ca\,II clouds are known to 
originate in high Galactic latitude (e.g.\ Sembach \& Danks 1994).  One 
possibility is that the Ca\,II absorber arises in a high velocity cloud in the 
halo of the DLA galaxy (Rauch \etal\ 2002).
 
\subsubsection{Kinematic Properties of The DLA}

  To measure the rotation curve of the absorbing galaxy toward Q0809$+$483 
along the stellar disk, we first analyzed the surface brightness profile 
observed in an HST/WFPC2 image taken with the F702W filter (Figure 1).  We 
found that the stellar disk can be characterized with an inclination angle $i 
= 48\pm 5$ degrees and a position angle of the major axis $\alpha = 82\pm 2$ 
degrees, measured north through east.  Next, we measured the observed velocity 
shear $v_{\rm obs}$ as a function of projected distance $\rho$ from the center 
of the galaxy, using the spatially and spectrally resolved [O\,II] emission 
doublet.  Following the steps described in \S\ 3.1.3, we measured a rotation 
curve of the galaxy deprojected to the stellar disk.  The results are presented
in Figure 8.  Due to the proximity of the galaxy to the QSO and the low surface
brightness of the emission-line feature, we were able to measure accurate 
velocity shear at three positions along the disk. 

  The velocity offset between the DLA and the systematic velocity of the 
absorbing galaxy places an additional point at large galactocentric radius.  
Figure 1 shows that the background QSO intersects the extended stellar disk at 
$\rho = 5.9\,h^{-1}$ kpc and $\phi = -54.6$ degrees, leading to $R = 7.4\pm 1.0
\,h^{-1}$ kpc and $v_{\rm DLA} = -320\pm 97$ \kms\ for the absorbing cloud.  
The errors include uncertainties in $i$ and $\alpha$ of the stellar disk.  The 
derived rotation velocity of the DLA is shown as the open circle in Figure 8.
Comparison of the rotation motion of the optical disk and the DLA shows that 
the DLA is moving along the direction expected from extrapolating the rotation
curve of the inner disk and suggests that the neutral gas extends to at least
$7\,h^{-1}$ kpc radius along the disk.

\subsection{The DLA at $z=0.524$ Toward AO\,0235$+$164}
 
  The DLA at $z=0.524$ toward AO\,0235$+$164 has been discussed in detail by
Chen \& Lanzetta (2003).  It was selected based on 21\,cm absorption and
confirmed to have $N(\hI) = (5\pm 1)\times 10^{21}$ \cmjj\ in subsequent HST 
spectroscopy using STIS (Junkkarinen \etal\ 2004).  An active galaxy was found
at the redshift of the DLA, $9.4\,h^{-1}$ kpc from the QSO line of sight 
(Burbidge \etal\ 1996), but high spatial resolution images from HST/WFPC2 also 
revealed complex morphology of the galaxy (Chen \& Lanzetta 2003).  Figure 1
shows two compact sources embedded in an extended nebular with the second 
source $4.8\,h^{-1}$ kpc from the QSO.

\subsubsection{Properties of The Neutral Gas}

  Additional QSO spectra obtained using STIS/CCD with the G430L and G750L 
gratings are available from the HST archive.  These data cover a spectral range
from 2900 \AA\ through 1 micron, allowing a detailed analysis of the metal 
content of the DLA.  

  Fe\,II and Mn\,II absorption features are seen in the G430L spectrum.  The
absorption-line profiles of five Fe\,II transitions and three Mn\,II 
transitions are presented in Figure 9.  We performed a simultaneous Voigt
profile analysis to all the observed features, assuming that Fe\,II and Mn\,II
originate in the same neutral gas region.  We derived a best-fit $\log\,N({\rm 
Mn\,II})=13.8\pm 0.1$ for a velocity width of $b=41.3\pm 5.8\,{\rm km} {\rm 
s}^{-1}$ and $N({\rm Fe\,II})=15.3\pm 0.4$.  The errors represent the 
1-$\sigma$ confidence interval as derived from VPFIT.  The large best-fit $b$ 
again suggests that there are multiple absorbing components in the gaseous 
cloud that are not resolved in the low-resolution FOS data.  The best-fit Voigt
profiles are shown in red curves in Figure 9 for comparison with the data.

  Combining the $N(\hI)$ and $N({\rm Fe\,II})$ measurements, we derive $[{\rm 
Fe}/{\rm H}]_{\rm obs} = -1.8\pm 0.4$ on the assumption that all iron is singly
ionized and in neutral gaseous clouds.  The observed Fe abundance indicates 
that the DLA has only $1/60$ solar metallicity, but Fe is known to be 
susceptible to dust depletion and a prominent dust absorption feature at 
rest-frame 2175 \AA\ has been observed in subsequent STIS spectroscopy.  
Junkkarinen \etal\ (2004) reported a color excess $E(B-V)=0.227\pm 0.003$ and a
general extinction coefficient $R_V=2.51\pm 0.03$ for the neutral gas.  
Following the prescription in Savaglio \& Fall (2004) and adopting the Galactic
dust-to-metal ratio, we derived a dust depletion correction of 1.2 dex for the 
Fe abundance, leading to a depletion-corrected metallicity of $[{\rm Fe}/{\rm 
H}]_{\rm undepleted} = -0.6\pm 0.4$ (or $1/4$ solar).

  Turnshek \etal\ (2003) have presented metallicity measurements for the DLA 
toward AO\,0235$+$164 based on an X-ray spectrum of the QSO.  To account for 
Galactic absorption in the X-ray data, these authors considered a range of 
Galactic gas metallicity models and derived a best-fit DLA metallicity of
0.24--0.64 solar.  Our depletion-corrected measurement agrees well with the
result of Turnshek \etal\ (2003) for the solar gas metallicity model.

\subsubsection{ISM Properties of The Absorbing Galaxy}

  The galaxy environment of the DLA is fairly complex.  In addition to the 
active galaxy at $2''$ angular distance to the background QSO, a compact source
at $1''$ angular distance away from the QSO and with $1/3$ of the luminosity of
the active galaxy (corresponding to $M_{AB}(B)-5\log h=-19.1$), has been 
identified at the redshift of the DLA in a deep HST/WFPC2 image (Figure 1).  

  The summed, extracted spectrum presented in Figure 10 displays multiple 
emission features that suggest a starburst or AGN environment.  At the same 
time, it also displays a strong 4000-\AA\ flux discontinuity, indicating an 
underlying old stellar population.  The spectrum covers 3400--5300 \AA\ in the 
rest frame of the galaxy and shows [Ne\,V], [O\,II], [O\,III], and H$\beta$ 
emission features.  Line flux measurements of the identified emission features 
are presented in Table 7.  The redshift of the galaxy is $z=0.52530\pm 
0.00007$, $\approx 200$ \kms\ away from the redshift of the DLA.

  We were unable to obtain an accurate estimate of dust extinction in the ISM 
of the galaxy due to the lack of spectral coverage over the H$\alpha$ 
transition, but the observed H$\gamma$/H$\beta$ ratio suggests little/no dust
extinction in galaxy.  We continued our analysis, assuming no extinction 
correction.  While the presence of [N\,V]\,$\lambda$3425 requires a hard 
ionizing source and indicates the presence of an active nucleus, the observed 
strong line ratios $\log$ [O\,III]\,$\lambda$5007/H$\beta = 0.53 \pm 0.03$ and
$\log$ [O\,II]\,$\lambda$3727/H$\beta = 0.37 \pm 0.03$ also suggest star 
formation not AGN activity as the dominant source for producing these spectral 
features (Rola, Terlevich, \& Terlevich 1997).  Our spectrum is therefore a
blend of both components.

  Assuming the strong lines largely originate in star forming regions, we 
estimated the oxygen abundance of the emission-line region using the $R_{23}$ 
index.  We measured $R_{23}=0.85\pm 0.03$, corresponding to $12 + \log({\rm 
O}/{\rm H}) = 8.5$ or 6.7 according to the analytic formula in Kobulnicky 
\etal\ (1999).  The double values in the predicted oxygen abundance differ by 
nearly two orders of magnitude.  We found the higher value to be more 
plausible, taking into account the luminosity-metallicity correlation observed 
in field emission-line galaxies (e.g.\ Kobulnicky \& Zaritsky 1999), and 
concluded that the ISM of the galaxy has $\approx 60$\% solar metallicity.  The
observed ISM metallicity appears to be 0.45 dex higher than the metallicity 
estimated for the neutral gas at $\rho=5-9 \ h^{-1}$ kpc, showing a trend of 
lower metallicity at larger galactic radii but with a large uncertainty.

\subsection{The DLA at $z=0.525$ Toward B2\,0827$+$243}

  The strong metal-line absorber at $z=0.525$ toward B2\,0827$+$243 has been
selected by the presence of a strong Mg\,II absorption doublet and confirmed to
have $N(\hI)=2\times 10^{20}$ \cmjj\ in subsequent HST observations using 
FOS (Rao \& Turnshek 2000).  An edge-on disk galaxy has been identified at the 
redshift of the DLA, $26.7\ h^{-1}$ kpc projected distance away from the QSO 
line of sight (Steidel \etal\ 2002).

\subsubsection{Properties of The Neutral Gas}

  The DLA exhibits a saturated Mg\,II $\lambda\lambda\,2796,2803$ doublet at $z
= 0.52499$ with a rest-frame absorption equivalent width of 2.47 \AA, among the
strongest Mg\,II systems known (Steidel \etal\ 2002).  A recent study of this
sightline by Khare \etal\ (2004) presents observations of a series of strong
Fe\,II and Mn\,II transitions of the DLA.  These metal lines are resolved into
two primary components in the moderate-resolution spectrum, separated by 50 
\kms.  The estimated ionic column densities of the two components together are 
$\log N({\rm Fe\,II})=14.74\pm 0.04$ and 
$\log N({\rm Mn\,II})=12.83^{+0.19}_{-0.34}$.  We combined the $N(\hI)$ and 
ionic column densities to derive $[{\rm Fe}/{\rm H}] = -1.01\pm 0.11$ and
$[{\rm Mn}/{\rm H}] = -0.86\pm 0.35$, assuming a 1-$\sigma$ error of 0.1 dex
in $N(\hI)$.

  Zinc is not observed in the moderate-resolution QSO spectrum, which 
prohibited us from obtaining an empirical estimate of dust depletion in the 
neutral gas.  Nevertheless, the abundance measurements show that $[{\rm 
Mn}/{\rm Fe}]=0.15\pm 0.37$---consistent with solar, while Mn is frequently 
observed to be underabundant with respect to Fe in DLAs (e.g.\ Pettini \etal\ 
2000; Dessauges-Zavadsky, Prochaska, \& D'Odorico 2002).  The observed 
$[{\rm Mn}/{\rm Fe}]$ is therefore suggestive of some dust depletion.  We again
adopted the median $[{\rm Zn}/{\rm Fe}]$ and the observed scatter (see \S\ 
3.1.1) to correct for dust depletion, and derived a metallicity $Z=-0.49 
\pm 0.30$ for the DLA at $z=0.5249$ toward B2\,0827$+$243.

\subsubsection{ISM Properties of The Absorbing Galaxy}

  The summed, extracted spectrum of the DLA galaxy covers 3400--5400 \AA\ and 
is presented in Figure 11.  The galaxy displays multiple emission features such
as [O\,II] and  H$\beta$, as well as a moderate 4000-\AA\ spectral 
discontinuity.  These spectral features indicate a combination of an evolved 
stellar population and ongoing star formation in the ISM of the DLA galaxy.  
Line flux measurements of the identified emission features are presented in 
Table 8.  [O\,III] $\lambda\,4959$ emission is not detected, so we present a 
3-$\sigma$ upper limit (over a resolution element $\approx 3.6$ \AA) to the 
line flux.  At the same time, the [O\,III] $\lambda\,5007$ emission feature 
appear to be uncharacteristically strong.  But because this feature coincides 
with the telluric A-band absorption, the observed emission-line flux is subject
to a large systematic uncertainty after correcting for the atmosphere 
absorption.  The line luminosity of [O\,III] $\lambda\,5007$ in Table 8 is 
therefore derived based on the expected 1:3 line flux ratio between the two 
[O\,III] forbidden transitions and the observed limit to the [O\,III] 
$\lambda\,4959$ line luminosity.  The redshift of the galaxy is $z =0.52631\pm 
0.00007$.

  We were unable to assess the dust content because of a lack of spectral 
coverage of the H$\alpha$ emission line.  In the absence of dust obscuration, 
we derived an upper limit to the $R_{23}$ metallicity index and found 
$\log\,R_{23} \le 0.54$.  Given the luminous nature of the galaxy
($M_{AB}(B)-5\log h = -20.0$), we again adopted the upper branch of the 
metallicity-$R_{23}$ correlation and derived a lower limit to the oxygen 
abundance $12 + \log({\rm O}/{\rm H}) \ge 8.8$ which is more than the typical
solar value.  This galaxy is the third system in our sample that shows a higher
metallicity in the inner disk ISM than one estimated in the neutral gas in the 
outskirt of the gaseous disk.

\subsubsection{Kinematic Properties of The DLA}

  The kinematic properties of this DLA galaxy have been studied by Steidel
\etal\ (2002).  They presented a rotation curve of the edge-on optical disk and
compared with the associated Mg\,II absorption profile observed in echelle 
mode.  The velocity offset between the DLA and the systematic velocity of the 
absorbing galaxy  places an additional point at large galactocentric radius.  
Figure 12 shows that the background QSO intersects the extended optical disk at
$\rho = 26.7\,h^{-1}$ kpc with a lag in velocity $v_{\rm DLA} = -191\pm 33$ 
\kms\ for the absorbing cloud.  No correction has been made for the disk 
orientation because of its edge-on nature and because the QSO sightline 
coincides with the extension of the edge-on disk (see Figure 1)  The derived 
rotation velocity of the DLA is shown as the open star in Figure 12.
Comparison of the rotation motion of the stellar disk and the DLA shows that 
the DLA is moving along with the optical disk and suggests that the neutral gas
extends to at least $25\,h^{-1}$ kpc radius along the disk.

\subsection{The DLA at $z=0.612$ Toward LBQS0058$+$0155}

  The DLA at $z=0.6118$ toward LBQS0058$+$0155 was first discovered by Pettini
\etal\ (2000) with $N(\hI) = (1.2 \pm 0.5) \times 10^{20}$ \cmjj\ in an HST/FOS
spectrum.  A detailed study of chemical abundances of the absorber by these 
authors showed that the absorbing gas has near-solar metallicity, $[{\rm Zn}/
{\rm H}]=0.08\pm 0.21$.  A candidate absorbing galaxy at $1.2''$ angular 
distance away from the QSO was also found in a deep HST/WFPC2 image of the 
field (Figure 1).  

  The summed, extracted spectrum of the candidate galaxy, covering a rest-frame
wavelength range from 3500 \AA\ through 4450 \AA\ is presented in Figure 13.  
Only a weak emission feature is identified in the spectrum, which is evident in
the two-dimensional spectral image presented in the inset of Figure 13.  
Identifying the line as [O\,II] led to an estimate of $z = 0.6120\pm 0.0002$, 
$\approx 40$ \kms\ away from the redshift of the DLA.  The galaxy is confirmed
to be responsible for the DLA.  At this redshift, the galaxy is $\rho =
5.5\,h^{-1}$ kpc from the QSO line of sight and has $M_{\rm AB}(B) - 5\,\log\,h
= -17.6$.

\section{DISCUSSION}

  We have obtained and analyzed optical spectra of six DLA galaxies at $z\apl
0.6$.  These galaxies exhibit a range of ISM properties, from post-starburst
(Q0738$+$313) through normal disks (PKS0439$-$433, Q0809$+$483, and 
B2\,0827$+$243) and through starburst (AO\,0235$+$164).  Their ISM properties 
support the idea that DLA galaxies are drawn from the typical field 
population, and not from a separate population of low surface brightness or 
dwarf galaxies.  A summary of measurable physical parameters of the absorbing 
galaxies is listed in Table 10, where we present from columns (2) through (8) 
the galaxy redshift $z_{\rm gal}$, projected distance $\rho$, absolute $B$-band
magnitude $M_{\rm AB}(B)$, rotation velocity $V_{\rm term}$, dynamic mass 
$M_{\rm dyn}$, and oxygen abundance $[{\rm O}/{\rm H}]$. 

  We have also analyzed ultraviolet spectra of the DLA systems toward 
PKS0439$-$433 and AO\,0235$+$164 for estimating metallicity in the neutral gas
clouds.  Supplementing our own measurements with available metallicity 
measurements in the literature, we have collected six DLA systems at $z<1$, 
for which metal abundances of the gaseous clouds are measured and can be 
compared directly with the physical properties of the absorbing galaxies.  A 
summary of measurable parameters of the neutral gas is also listed in Table 10,
where we present from columns (9) through (12) the absorber redshift $z_{\rm 
DLA}$, column density measurements and associated errors for \hI\ and Fe\,II, 
and dust-depletion corrected metallicity in the neutral gas $Z$\footnote{The
gas metallicity is defined as $Z\equiv\log({\rm X}/{\rm H})-\log({\rm X}/{\rm 
H})_\odot$, where X represents dust depletion-corrected metal abundances.}.  We
have included the DLA at $z_{\rm DLA}=0.6819$ toward HE1122$-$1649 and the DLA 
at $z_{\rm DLA}=1.0095$ toward EX0302$-$2223 in Table 10.  Photometric 
properties of the absorbing galaxies are measured by Chen \& Lanzetta (2003). 
Metal abundances of the absorbers are measured by Ledoux, Bergeron, \& 
Petitjean (2002) for the DLA system toward HE1122$-$1649 and by Pettini \etal\ 
(2000) for the DLA system toward EX0302$-$223.

  The galaxy sample presented in Table 10 can be adopted to address two 
important issues.  In this section, we discuss in turn the Tully-Fisher 
relation of $N(\hI)$-selected galaxies at intermediate redshifts and abundances
profiles of the galaxies.  In particular, we investigate whether or not the low
metal content of the DLA population arises naturally as a combination of gas 
cross-section selection and metallicity gradients.

\subsection{The Tully-Fisher Relation}

  Four galaxies in the DLA galaxy sample presented here appear to have disk 
morphology (Figure 1).  We have successfully extracted a rotation curve 
measurement along the major axis of the optical disk of the galaxies toward 
PKS0439$-$433 (Figure 5), Q0809$+$483 (Figure 8), and B2\,0827$+$243 (Figure 
12).  We note that the observed inner slope of these rotation curves is 
expected to be shallower than the intrinsic shape because of a 
``beam-smearing'' effect caused by seeings that has not been removed from our
data.  Nevertheless, the observations allow us to compare the relative motions 
between the hot gas in the inner ISM and the neutral gaseous clouds at large
galactocentric distances, and to obtain a robust measurement of the rotation 
speed of galaxy disks at intermediate redshifts.  In all three cases, we 
observe consistent rotation motion between the optical disks and the neutral 
gaseous clouds.  The results of our rotation curve study supports that galaxies
with disk-like stellar morphologies at intermediate redshifts possess large 
rotating \hI\ disk out to $30\ h^{-1}$ kpc, comparable to what is observed in 
local disk galaxies.

  Our small DLA galaxy sample serves as raw material for a pilot study of the 
Tully-Fisher relation  (Tully \& Fisher 1977) of intermediate-redshift 
galaxies.  We have estimated the terminal velocity $V_{\rm term}$ of each disk 
based on a comparison of the redshifts of the absorber and galaxy, corrected 
for the inclination of the disk and the relative orientation of the QSO 
sightline along the disk (see \S\ 3.1.3).  Figures 5, 8, and 12 together show 
that the absorbers offer additional constraints in the disk rotation speed at 
distances that are more than twice beyond the extent of the optical disks.  In 
particular, the rotation curves for the DLA galaxies toward PKS0439$-$433 and 
B2\,0827$+$243 remain flat beyond the optical disks.  We estimate the enclosed 
dynamic mass $M_{\rm dyn}$ of each absorbing galaxy based on the terminal 
velocity measured at the location of the absorber ($R=9.5\ h^{-1}$ kpc for the 
DLA toward PKS0439$-$433; $R=7.4\ h^{-1}$ kpc for the DLA toward Q0809$+$483; 
$R=26.7\ h^{-1}$ kpc for the DLA toward B2\,0827$+$243).  The results are 
presented in column (7) of Table 10.  Following $M_{\rm dyn} = R\,v_{\rm 
term}^2\,/\,G$, we find that these three DLA systems reside in massive halos of
$>10^{11}\,h^{-1}\,{\rm M}_\odot$, comparable to those of Milky-Way type 
galaxies.  We note that the three examples presented here are among the 
brightest of known DLA galaxies at $z<1$, and our results do not exclude the 
possibility that some DLA systems reside in halos of smaller mass scales.

  Figure 14 shows the Tully-Fisher relation of the DLA galaxies (open stars), 
in comparison to the best-fit $B$-band relation derived from \hI\ 21\,cm 
velocity measurements of local galaxies (Pierce \& Tully 1988, 1992; the solid 
line) and to a field sample of intermediate-redshift galaxies (Vogt \etal\ 
1997; the dotted line).  While the field galaxy sample---selected based on 
optical brightness---exhibits a moderate luminosity brightening of $0.36\pm 
0.13$ mag (Vogt \etal\ 1997), the DLA galaxy sample---selected based on 
$N(\hI)$---shows a brightening of $0.2\pm 0.5$ mag for a fixed slope determined
in the local $B$-band relation, consistent with little/no evolution in the 
Tully-Fisher relation at similar redshift range.  

\subsection{Abundance Profiles}

  To investigate whether or not metallicity gradients exist in 
intermediate-redshift galaxies, we first compare abundance measurements of the
ISM and the neutral gas in individual systems.  Figure 15 shows the observed 
metallicities in the inner ISM of the galaxy and in the neutral gas at large 
galactic radius $R$ for three DLA systems in our sample (PKS0439$-$433, 
AO\,0235$+$164, and B2\,0827$+$243).  In each pair, we have adopted the oxygen 
abundance measured in the gas phase of the inner ISM and the Fe abundance for 
the neutral gas, corrected for dust depletion.  All abundance measurements are 
normalized to their respective solar values.  Note that the DLA toward 
AO\,0235$+$164 is presented at its projected distance to the absorbing galaxy, 
because the galaxy exhibits complex morphology that cannot be represented by a 
simple disk model.  

  We find in all three cases an abundance decrement by a factor of $3-5$ (or
0.45 -- 0.7 dex) from the center of the galaxies to neutral gaseous clouds at 
$9-26\ h^{-1}$ kpc galactic radii, corresponding to a radial gradient ranging 
from $-0.02$ to $-0.07$ dex kpc$^{-1}$.  These values are consistent with the 
range of abundance gradients observed in local disk galaxies (Zaritsky \etal\ 
1994).  We note that although the ISM abundances are derived based on strong 
emission lines characteristic of star forming H\,II regions, the emission-line 
based abundance measurements are representative of the mean metal content of 
the local atomic and molecular phases as well\footnote{See Kennicutt \etal\ 
(2003) for a comparison of H\,II region abundances from Deharveng \etal\ (2000)
and diffuse ISM abundances from Meyer, Jura, \& Cardelli (1998) and Moos 
\etal\ (2002) in the Galaxy.}.  The results in Figure 15 suggest that these 
three DLA systems originate in the relatively unevolved outskirts of galactic 
disks.  



  There are, however, systematic uncertainties that may reduce the gradients
seen in the three DLA galaxies presented in Figure 15.  First, the strong-line 
``empirical" H\,II region abundance scale has been called into question 
recently, because direct abundance measurements based on electron temperature 
($T_{\rm e}$) sensitive auroral lines tend to give abundances that are 
systematically lower by 0.2--0.5 dex (e.g.\ Kennicutt \etal\ 2003; Bresolin 
\etal\ 2004).  Such uncertainty has an impact on the interpretation of Figure 
15,
where the strong-line H\,II region abundances are being compared to absorption 
line abundances using completely different methods.  If we would lower the 
H\,II region abundances shown in Figure 15 by 0.2--0.5 dex, the radial 
gradients would be reduced substantially in at least two of the three systems 
(B2\,0827$+$243 and AO\,0235$+$164).  But it is unclear whether this correction
should be applied in full.  Recombination line abundances for Galactic H\,II 
regions yield results that are intermediate between the $T_{\rm e}$-based and 
strong-line values (e.g.\ Esteban \etal\ 1998).  Moreover, we have not included
a dust depletion correction in the oxygen abundance measurements for these two
systems.  Including depletion correction would increase the abundances by 
$\approx 0.1$ dex (e.g.\ Esteban \etal\ 1998).

  Second, the fraction of Fe locked in dust is observed to be higher at higher
metallicity regime, although there is a large scatter about the mean relation 
(e.g.\ Pettini \etal\ 1997; Prochaska \etal\ 2002).  We have adopted a median 
[Zn/Fe] ratio to estimate a depletion correction for the DLA systems toward 
PKS0439$-$433 and B2\,0827$+$243.  The correction may be underestimated 
particularly for the DLA system toward PKS0439$-$433, for which we have derived
a dust-depletion corrected metallicity of nearly solar, $[{\rm Fe}/{\rm H}] = 
-0.14 \pm 0.30$.  An underestimated Fe abundance would also reduce the 
abundance gradient observed in Figure 15 for the DLA system.  To assess the 
systematic uncertainty due to metallicity-dependent dust-depletion correction, 
we adopt the empirical relation between depletion fraction $f_{\rm Fe}$ and 
metallicity $[{\rm Fe}/{\rm H}]$ from Vladilo (2004) based on a study of 34 DLA
systems that have observed Fe abundance ranging from $[{\rm Fe}/{\rm H}]_{\rm 
obs}=-2$ to $-0.3$ and dust-depletion corrected Fe abundance ranging from 
$[{\rm Fe}/{\rm H}] = -2$ to $0.4$.  We derive a dust-depletion corrected 
$[{\rm Fe}/{\rm H}] = -0.24$ to $1.23$ for the DLA system toward PKS0439$-$433,
but we note that the empirical relation of dust depletion correction by Vladilo
is not well constrained by the data beyond $[{\rm Fe}/{\rm H}]=0$.  Comparing 
with the high-metallicity DLA systems in Vladilo's sample, we find that the DLA
system toward PKS0439$-$433 is likely to have a dust-depletion corrected 
metallicity of between $[{\rm Fe}/{\rm H}] = -0.24$ and 0.26, consistent with 
our previous estimate using a median [Zn/Fe] ratio to within errors.  We 
therefore conclude that the systematic uncertainty due to metallicity-dependent
dust-depletion correction is insignificant in the absorption-line abundance 
measurement for this DLA system.
 
  Finally, the abundance gradients observed in Figure 15 is based on 
comparisons between oxygen and iron abundances with respect to the typical 
solar values, but it is not clear whether the (O/Fe) ratio is solar in the DLA 
systems. An abundance pattern with enhanced Fe would steepend the abundance 
gradients, and enhanced oxygen would reduce the gradients.  While DLA systems 
show a large scatter in the observed $[{\rm O}/{\rm Fe}]$ around $[{\rm O}/{\rm
Fe}]=0$ (Lu \etal\ 1996; Prochaska, Howk, \& Wolfe 2003; D'Odorico \& Molaro 
2004), dwarf stars in the Galaxy show $[{\rm O}/{\rm Fe}] < 0.2$ at $[{\rm 
Fe}/{\rm H}] > -0.6$ and reach $[{\rm O}/{\rm Fe}] \approx 0$ at $[{\rm 
Fe}/{\rm H}] = 0$ (McWilliam 1997).  We therefore conclude that a solar (O/Fe) 
ratio remains as a reasonable approximation for the abundance comparison in 
Figure 15.

  To minimize the systematic uncertainties discussed above, we study 
metallicity gradients by considering only absorption-line abundances.  We 
investigate whether there exists a statistical correlation between 
absorption-line metallicity and galaxy impact parameter using an ensemble of 
six galaxy-DLA pairs for which Fe abundance measurements in the absorbers are 
available.  Given a mix of galaxy morphology, here we consider only the 
projected distances between the absorbers and the absorbing galaxies, rather 
than galactocentric radii.  The goal is to establish a mean profile that
characterizes the global distribution of metals in individual galaxies.  
Figure 16 shows the metallicity of neutral gas $Z$ versus galaxy impact 
parameter $\rho$ for the five galaxy-DLA pairs.  All measurements have been
corrected for dust depletion.  We find a declining trend of metallicities in 
neutral gaseous clouds with increasing impact parameter (open stars in Figure 
16).  The data are best described by a simple correlation,
\begin{equation}
Z(\rho)=(0.02\pm 0.20) - (0.041\pm 0.012) \frac{\rho}{({\rm kpc})},
\end{equation}
which is represented by the solid line in Figure 16.  The best-fit slope
$-(0.041\pm 0.012)$ corresponds to a scale length of $10.6\ h^{-1}$ kpc.  For 
comparison, we have also included in the figure the observed oxygen abundance 
gradient of M101 (Kennicutt \etal\ 2003) and the metallicity of $z<0.5$ 
intergalactic medium (IGM) from Prochaska \etal\ (2004).  The IGM metallicity 
has been estimated based on five \lya\ absorbers of $N(\hI)=10^{14-16}$ \cmjj\ 
at $z=0.09-0.5$ and is placed at impact parameter beyond $\rho=100\ h^{-1}$ kpc
according to the $N(\hI)$ versus $\rho$ correlation from Chen \etal\ (2001).

  Equation (4) shows an anti-correlation between metallicities in the cold gas 
and galaxy impact parameter at the 3-$\sigma$ level of significance, which 
applies to galaxy-DLA pairs within an impact parameter range of $\rho= 0 - 30
\ h^{-1}$ kpc.  The slope of the abundance profile agrees well with the oxygen 
abundance gradient observed in the H\,II region of M101, although flat 
metallicity distributions are also observed in some nearby disk galaxies 
(e.g.\ Zaritsky \etal\ 1994).  The agreement supports the idea that the DLA 
population traces galactic disks at intermediate redshifts.  Figure 16 also 
suggests that at $\rho=30\ h^{-1}$ kpc the neutral gas phase metallicity may 
already be approaching the IGM abundance floor at $z<0.5$.  We note that 
including intrinsic $B$-band luminosity of the absorbing galaxies does not 
improve the fit and shows little correlation between metallicity and galaxy 
luminosity using the small galaxy-absorber pair sample.

\subsection{Implications for the Mean Metallicity of the DLA Population}

  Consistent results have been presented by different authors over the past
few years to show that the mean metallicities observed in the DLA systems are 
only a tenth of the solar values at all redshifts that have been studied 
(e.g.\ Pettini \etal\ 1999; Prochaska \etal\ 2003a).  At the same time, some 
fraction of DLA systems are known to arise in disk galaxies at intermediate 
redshifts (e.g.\ Rao \etal\ 2003; Chen \& Lanzetta 2003; Lacy \etal\ 2003).  
While it is possible that the DLA systems are biased toward low-metallicity, 
low surface brightness galaxies (e.g.\ Schulte-Ladbeck \etal\ 2004), here we 
argue that the low metal content of the DLA population arises naturally as a 
combination of gas cross-section selection and metallicity gradients commonly 
observed in local disk galaxies (e.g.\ Zaritsky \etal\ 1994; Ferguson, 
Gallagher, \& Wyse 1998; van Zee \etal\ 1998).

  Adopting the radial gradient from equation (4), we proceed to construct a 
crude model for predicting the expected mean metallicity of the DLA population.
We consider first a uniform distribution of \hI\ gas.  The probability of 
finding a DLA system arising in the impact parameter interval $\rho$ and 
$\rho+\Delta\,\rho$ for a gaseous halo of size $R$ is 
\begin{equation}
d\,P = \frac{2\,\pi\rho\,\Delta\,\rho}{\pi\,R^2}.
\end{equation}
The expected mean metallicity is then defined as
\begin{equation}
\langle Z\rangle=\int_0^R Z(\rho)\times \frac{2\,\rho}{R^2}\,d\,\rho.
\end{equation}
Applying equation (4) to equation (6) yields a gas cross-section weighted mean 
metallicity $\langle Z\rangle=-0.64\pm 0.28$ for $R=24\ h^{-1}$ kpc, which is 
equivalent to the expected mean metallicity over an ensemble of DLA systems 
under the assumption that all DLA galaxies share this global abundance profile.
Our prediction is consistent with observations to within errors at $z\sim 1$, 
where the reported mean metallicity is $\langle Z \rangle_{\rm DLA}=-1.03\pm 
0.34$ (Prochaska \etal\ 2003a).

  Next, we consider empirical H\,I profiles of local disk galaxies for
calculating a $N(\hI)$-weighted mean metallicity of the DLA population.
Lanzetta, Wolfe, \& Turnshek (1995) have demonstrated that the cosmic 
metallicity associated with the DLA systems can be derived from calculating the
column density weighted average of the metallicities of the individual 
absorbers.  We perform a Monte-Carlo analysis for predicting the expected 
$N(\hI)$-weighted mean metallicity.  First, we adopt the mean H\,I profiles of 
Sab, Sbc, and Scd galaxies from Cayatte \etal\ (1994) for the field galaxy 
sample published in Broeils \& van Woerden (1994).  Next, we form a sample of 
100 DLA systems randomly drawn from different locations of the adopted H\,I 
disk profile, including the scatter in the mean profile ($\Delta\log N(\hI)
\approx 0.2$) and the gas cross-section selection effect.  Next, we assign a 
neutral gas metallicity at the location of each DLA systems, according to 
equation (4) with uncertainties included.  Finally, we sum together all the 
metals found in the 100 absorbers and derive the expected $N(\hI)$-weighted 
mean metallicity.

  The results of our Monte-Carlo simulations are presented in Figure 17 for
three different types of galaxies.  The top panels show for each galaxy type 
the $N(\hI)$ distribution versus radius in the simulated sample of DLA systems.
The bottom panels show their simulated metallicity distribution versus radius.
In the lower-left corner of the bottom panels, we list mean metallicity 
$\langle Z\rangle$ and $N(\hI)$-weighted mean metallicity
$\langle Z\rangle_{N(\hI)-{\rm weighted}}$ for each simulated DLA sample.  In
1000 sets of simulations, we find a $N(\hI)$-weighted mean metallicity $\langle
Z\rangle_{N(\hI)-{\rm weighted}}=-0.50\pm 0.07$ over 100 random sightlines,
where the error marks the 95\% uncertainty interval.  The is in accord with 
$\langle Z\rangle_{{\rm DLA}; N(\hI)-{\rm weighted}} = -0.64_{-0.86}^{+0.40}$ 
from Prochaska \etal\ (2003a), but is marginally higher than $\langle Z
\rangle_{{\rm DLA}; N(\hI)-{\rm weighted}} = -0.94_\pm 0.32$, $2 \sigma$ 
errors, from Kulkarni \etal\ (2004).  

  The difference between the two sets of $N(\hI)$-weighted mean metallicity 
measurements from Prochaska \etal\ and Kulkarni \etal\ arises mainly as a 
result of whether the DLA system toward AO\,0235$+$164 is included in the
analysis.  While x-ray absorption based metallicity measurement depends 
sensitively on how well the foreground absorption can be constrained, the
measurement from Turnshek \etal\ (2003) agrees well with our UV-absorption line
based measurement $[{\rm Fe}/{\rm H}]=-0.6\pm 0.4$.  The agreement therefore
justifies the inclusion of the absorber toward AO\,0235$+$164 in the 
$N(\hI)$-weighted mean metallicity calculation by Prochaska \etal.  The 
discrepancy resulted from inclusion or exclusion a single system in calculating
the mean metallicity, however, signifies the large uncertainty associated with
working with a small sample.  We argue based on the Monte-Carlo simulations 
that to obtain a representative estimate of the cosmic mean metallicity in the 
neutral gas phase requires a sample of DLA systems collected from $\sim 60$ 
random lines of sight.

  Finally, we draw attention to the scatter in metallicity displayed in Figure
17.  Pettini \etal\ (1994) performed a similar simulation to study whether
metallicity gradients could explain the large scatter observed among the
abundance measurements of individual DLA systems at high redshift.  These 
authors adopted oxygen abundance profiles of 13 nearby galaxies for which \hI\ 
profiles are available.  In all but one galaxy (M101), they were unable to 
reproduce the large observed abundance scatter.  Our simulations show that over
100 random sightlines drawn from within a $R=24\ h^{-1}$ kpc radius, we expect 
a 1-$\sigma$ (68\%) scatter of 0.7 dex, comparable to what is observed in the 
DLA systems at $z\sim 2$ (Prochaska \etal\ 2003a).  The fact that we are able 
to reproduce the large scatter in our simulations adds further support for the 
scenario in which the low metallicities observed in the $z<1$ DLA systems arise
naturally as a combination of gas cross-section and metallicity gradients.  
But whether or not the same scenario can be applied to explain the low metal 
content of the DLA systems at $z\sim 3$ is not clear, because large gaseous 
disks of $24\ h^{-1}$ kpc radius may not exist at this early epoch.

  In summary, our analysis demonstrates that metallicity gradients together 
with a gas cross-section selection can explain the low metallicities observed 
in the DLA systems at $z<1$.  It confirms that the low metal content of DLA 
systems does not rule out the possibility that the DLA population trace the 
field galaxy population.  Whether or not the same scenario can be applied to 
the DLA systems at higher redshifts remains to be verified.

\section{SUMMARY AND CONCLUSIONS}

  We have completed a spectroscopic study of six DLA systems at $z<0.65$, based
on moderate-to-high resolution spectra of the galaxies responsible for the 
absorbers.  We present two new identifications of DLA galaxies (Q0809$+$483 at 
$z_{\rm DLA}=0.437$ and LBQS0058$+$0155 at $z_{\rm DLA}=0.613$) and four DLA 
galaxies that are known previously (PKS0439$-$433 at $z_{\rm DLA} = 0.101$, 
Q0738$+$313 at $z_{\rm DLA}=0.221$, AO\,0235$+$164 at $z_{\rm DLA} = 0.524$, 
and B2\,0827$+$243 at $z_{\rm DLA}=0.525$).  We also present two new 
metallicity measurements of the absorbers for the DLA systems toward
PKS0439$-$433 and AO\,0235$+$164 based on the ultraviolet spectra of the 
background QSOs retrieved from the Hubble Space Telescope (HST) data archive.

  Combining known metallicity measurements of the absorbers with known optical
properties of the absorbing galaxies, we investigate whether or not the low 
metal content of the DLA population arises naturally as a combination of gas 
cross-section selection and metallicity gradients commonly observed in local 
disk galaxies.  Furthermore, we study disk evolution based on a comparison of 
the Tully-Fisher relation of the DLA galaxies and those of field galaxies.
The results of our study are summarized as follows:

  1. The galaxies responsible for DLA systems exhibit a range of ISM 
properties, from post-starburst, to normal disks, and to starburst systems.
Their ISM properties support the idea that DLA galaxies are drawn from the
typical field population, and not from a separate population of low surface
brightness or dwarf galaxies.

  2. The low metal content of the $z<1$ DLA population arises naturally as a 
combination of gas cross-section weighted selection (which favors large radii) 
and a metallicity gradient.

  3. Large rotating \hI\ disks of radius $30\ h^{-1}$ kpc are common at 
intermediate redshifts.  In addition, a rotation-curve study of three DLA 
galaxies shows that they reside in massive halos of $>10^{11}\,h^{-1}\,{\rm 
M}_\odot$, comparable to those of Milky-Way type galaxies rather than dwarfs

  4. A comparison of the Tully-Fisher relation of three DLA galaxies and those
of emission-line selected field samples shows little detectable evidence for 
evolution in the disk population between $z=0$ and $z\sim 0.5$.

  5. We observe an abundance decrement by a factor of $3-5$ from the center of 
the galaxies to $9-26\ h^{-1}$ kpc galactocentric radius away for all three DLA systems that have abundance measurements both in the inner ISM of the galaxy 
(derived using the $R_{23}$ metallicity index) and in the neutral gas (derived 
according to a dust-depletion corrected Fe abundance).  But including 
uncertainties in the empirical calibration of the $R_{23}$ index may reduce
the abundance decrement for two of the three cases.

  6. Using an ensemble of six galaxy-DLA pairs, we derive an abundance profile
that can be characterized by a radial gradient of $-0.041\pm 0.012$ dex per 
kiloparsec over an impact parameter range of $\rho=0-30\ h^{-1}$ kpc or
equivalently a scale length of $10.6\ h^{-1}$ kpc.

  7. Adopting known $N(\hI)$ profiles of nearby galaxies and the best-fit 
radial gradient, we further derive an $N(\hI)$-weighted mean metallicity 
$\langle Z\rangle_{\rm weighted} = -0.50\pm 0.07$ over 100 random lines of
sight for the DLA population, consistent with $\langle Z\rangle_{\rm weighted} 
= -0.64_{-0.86}^{+0.40}$ observed for $z\sim 1$ DLA systems.

  8. Monte-Carlo simulation results indicate that a sample of $\sim 60$ DLA 
systems drawn from random lines of sight is required for accurate estimates of
the cosmic mean metallicity in the neutral gas phase.

\acknowledgments   
We appreciate the expert assistance from the staff of the Las Campanas 
Observatory and the MMT Observatory.  It is a pleasure to thank Scott Burles, 
Jason Prochaska and Sandra Savaglio for interesting discussions, and Paul 
Schechter and Rob Simcoe for helpful comments on an earlier draft.  We thank
the referee, Max Pettini, for a speedy review and insightful comments that 
helped to improve the presentation of the paper.  H.-W.C. acknowledges support 
by NASA through a Hubble Fellowship grant HF-01147.01A from the Space Telescope
Science Institute, which is operated by the Association of Universities for 
Research in Astronomy, Incorporated, under NASA contract NAS5-26555.  
R.\ C.\ K.\ acknowledges the support of the NSF through grant AST 98-11789 and 
NASA through grant NAG5-8426.  M.\ R.\ is grateful to the NSF for grant AST 
00-98492 and to NASA for grant AR 90213.01-A.

\newpage

\tiny
\begin{deluxetable}{p{1.3in}ccccccrc}
\rotate
\tablecaption{Summary of the DLA galaxies}
\tablewidth{0pt}
\tablehead{\colhead{} & \colhead{} & \colhead{} & \colhead{$\log N(\hI)$} & 
\colhead{$\Delta\alpha$} & \colhead{$\Delta\delta$} & \colhead{$\Delta\theta$} 
& \colhead{} & \colhead{} \\
\colhead{Sightline} & \colhead{$z_{\rm QSO}$} & \colhead{$z_{\rm DLA}$} & 
\colhead{(cm$^{-2}$)} & \colhead{(arcsec)} & \colhead{(arcsec)} & 
\colhead{(arcsec)} & \colhead{$AB$} & \colhead{Morphology} \\
\colhead{(1)} & \colhead{(2)} & \colhead{(3)} & \colhead{(4)} & \colhead{(5)} &
\colhead{(6)} & \colhead{(7)} & \colhead{(8)} & \colhead{(9)} }
\startdata
PKS0439$-$433 \dotfill   & 0.593 & 0.101 & 20.0 & $-0.8$ & $+4.0$ & 4.1 & $I=17.2$ & disk \nl
Q0738$+$313 \dotfill     & 0.635 & 0.221 & 20.9 & $+2.0$ & $-5.0$ & 5.4 & $I=20.9$ & compact \nl
Q0809$+$483 \dotfill     & 0.871 & 0.437 & 20.8 & $+1.2$ & $-0.9$ & 1.5 & $I=19.9$ & disk \nl
AO\,0235$+$164 \dotfill    & 0.940 & 0.524 & 21.7 & $+0.7$ & $-1.1$ & 2.1 & $I=20.2$ & compact \nl
B2\,0827$+$243 \dotfill  & 0.939 & 0.525 & 20.3 & $+5.8$ & $+2.0$ & 6.1 & $R=21.0$ & disk \nl
LBQS0058$+$0155 \dotfill & 1.954 & 0.612 & 20.1 & $+0.8$ & $+0.9$ & 1.2 & $R=23.7$ & disk 
\enddata
\end{deluxetable}
\normalsize

\newpage

\begin{deluxetable}{p{1.3in}ccrcccc}
\rotate
\tablecaption{Journal of Ground-based Galaxy Spectroscopy}
\tablewidth{0pt}
\tablehead{\colhead{} & \colhead{} & \colhead{} & 
\colhead{} & \colhead{Blocking} & \colhead{} & \colhead{Spectral} & 
\colhead{Exposure} \\
\colhead{Field} & \colhead{Telescope} & \colhead{Instrument} & 
\colhead{Grating} & \colhead{Filter} & \colhead{$\Delta\lambda/{\rm pixel}$} &
\colhead{Coverage} & \colhead{Time (s)}}
\startdata
PKS0439$-$433 \dotfill  & Magellan/Clay & B\&C & 600 l/mm & GG400 & 1.6 \AA\ & 
3960 -- 7500 \AA\ & 1200 \nl
                        & Magellan/Clay & B\&C & 1200 l/mm & GG400 & 0.8 \AA\ &
4720 -- 6350 \AA\ & 2400 \nl
Q0738$+$313 \dotfill     & MMT & Blue Channel & 500 l/mm & LP495 & 1.2 \AA\ & 
3850 -- 8250 \AA\ & 2400 \nl
Q0809$+$483 \dotfill     & MMT & Blue Channel & 500 l/mm & LP495 & 1.2 \AA\ & 
5100 -- 8300 \AA\ & 4800 \nl
                         & MMT & Blue Channel & 1200 l/mm & L-42 & 0.5 \AA\ & 
4950 -- 6050 \AA\ & 3600 \nl
AO\,0235$+$164 \dotfill  & MMT & Blue Channel & 500 l/mm & LP495 & 1.2 \AA\ &
5100 -- 8300 \AA\ & 3600 \nl
B2\,0827$+$243 \dotfill  & MMT & Blue Channel & 500 l/mm & LP495 & 1.2 \AA\ &
5100 -- 8300 \AA\ & 3600 \nl
                         & MMT & Blue Channel & 1200 l/mm & L-42 & 0.5 \AA\ &
4950 -- 6050 \AA\ & 5400 \nl
LBQS0058$+$0155 \dotfill & Magellan/Clay & B\&C & 600 l/mm & OG570 & 1.6 \AA\ &
5600 -- 7200 \AA\ & 3600 \nl
\enddata
\end{deluxetable}

\newpage

\begin{deluxetable}{p{1.3in}ccrcccc}
\rotate
\tablecaption{Journal of HST Archival QSO Spectroscopy}
\tablewidth{0pt}
\tablehead{\colhead{} & \colhead{} & 
\colhead{} & \colhead{} & \colhead{} & \colhead{Spectral} & 
\colhead{Exposure} & \colhead{Proposal} \\
\colhead{Field} & \colhead{Instrument} & \colhead{Grating} & \colhead{Aperture}
& \colhead{$\Delta\lambda/{\rm pixel}$} & \colhead{Coverage} & \colhead{Time (s)} 
& \colhead{Identifier}}
\startdata
PKS0439$-$433 \dotfill  & STIS & G140L & $52\times 0.2$ & 1.6 \AA\ & 
1120 -- 1717 \AA\ & 2510 &  9382 \nl
                        & FOS & G190H & $0.25\times 2.0$ & 0.36 \AA\ & 
1610 -- 3277 \AA\ & 7888 &  4581 \nl
                        & FOS & G270H & $0.25\times 2.0$ & 0.51 \AA\ & 
1610 -- 3277 \AA\ & 2517 &  4581 \nl
AO\,0235$+$164 \dotfill  & STIS & G230L & $52\times 0.5$ & 1.4 \AA\ &
1585 -- 3141 \AA\ & 12830 & 7294 \nl
                         & STIS & G430L & $52\times 0.5$ & 2.6 \AA\ &
2900 -- 5710 \AA\ & 2874 & 7294 \nl
                         & STIS & G750L & $52\times 0.5$ & 4.9 \AA\ &
5270 -- 10260 \AA\ & 2160 & 7294 \nl
\enddata
\end{deluxetable}

\newpage

\begin{deluxetable}{p{0.6in}lrrrrr}
\tablecaption{Spectral features identified in the DLA galaxy at $z=0.10104$
toward PKS0439$-$433}
\tablewidth{0pt}
\tablehead{\colhead{Features} & \colhead{$\lambda_{\rm rest}$} &
\colhead{$\lambda_{\rm obs}$} &  
\colhead{Flux (${\rm erg}/{\rm sec}/{\rm cm}^2$)} & 
\colhead{${\rm EW}_{\rm rest}$ (\AA)} 
& \colhead{$L$ ($h^{-2}\,{\rm erg}/{\rm sec}$)} & \colhead{Corrected $L$}}
\startdata
$[{\rm O\,II}]$ \dotfill & 3728.23 & 4105.82 & $(3.9 \pm 0.2)\times 10^{-16}$ &
$5.3 \pm 1.3$ & $4.9 \times 10^{39}$ & $1.2 \times 10^{40}$ \nl
H\,$\beta$  \dotfill & 4862.70 & 5354.09 & $(4.44 \pm 0.06)\times 10^{-16}$ & 
$4.6 \pm 0.6$ & $5.6 \times 10^{39}$ & $1.1 \times 10^{40}$ \nl
$[{\rm O\,III}]$ \dotfill & 4960.29 & 5462.99 & $(7.6 \pm 0.4)\times 10^{-17}$ &
$0.9 \pm 0.3$ & $9.6 \times 10^{38}$ & $1.9 \times 10^{39}$ \nl
$[{\rm O\,III}]$ \dotfill & 5008.24 & 5514.01 & $(1.40 \pm 0.04)\times 10^{-16}$ &
$1.5 \pm 0.5$ & $1.8 \times 10^{39}$ & $3.6 \times 10^{39}$ \nl
$[{\rm N\,II}]$ \dotfill & 6549.91 & 7211.74 & $(1.81 \pm 0.07)\times 10^{-16}$ &
$2.3 \pm 0.6$ & $2.3 \times 10^{39}$ & $3.8 \times 10^{39}$ \nl
H\,$\alpha$ \dotfill & 6564.63 & 7227.83 & $(1.61 \pm 0.02)\times 10^{-15}$ &
$20.7 \pm 1.9$ & $2.0 \times 10^{40}$ & $3.3 \times 10^{40}$ \nl
$[{\rm N\,II}]$ \dotfill & 6585.42 & 7250.92 & $(6.55 \pm 0.13)\times 10^{-16}$ &
$8.4 \pm 1.2$ & $8.3 \times 10^{39}$& $1.4 \times 10^{40}$ \nl
$[{\rm S\,II}]$ \dotfill & 6718.95 & 7397.32 & $(2.81 \pm 0.08)\times 10^{-16}$ &
$3.4 \pm 0.7$ & $3.6 \times 10^{39}$ & $5.8 \times 10^{39}$ \nl
$[{\rm S\,II}]$ \dotfill & 6733.16 & 7412.48 & $(2.44 \pm 0.07)\times 10^{-16}$ &
$2.9 \pm 0.7$ & $3.1 \times 10^{39}$ & $5.0 \times 10^{39}$ 
\enddata
\end{deluxetable}

\begin{deluxetable}{p{0.6in}lrrrr}
\tablecaption{Spectral features identified in the DLA galaxy at $z=0.2222$
toward Q0738$+$313}
\tablewidth{0pt}
\tablehead{\colhead{Features} & \colhead{$\lambda_{\rm rest}$} &
\colhead{$\lambda_{\rm obs}$} &  
\colhead{Flux (${\rm erg}/{\rm sec}/{\rm cm}^2$)} & 
\colhead{${\rm EW}_{\rm rest}$ (\AA)} & 
\colhead{$L$ ($h^{-2}\,{\rm erg}/{\rm sec}$)}}
\startdata
$[{\rm O\,II}]$ \dotfill & 3728.23 & 4556.64 & $< 2.2 \times 10^{-18}$ &
$< 2.9$  & $< 6.9 \times 10^{37}$ 
\enddata
\end{deluxetable}

\begin{deluxetable}{p{0.6in}lrrrr}
\tablecaption{Spectral features identified in the DLA galaxy at $z=0.43745$
toward Q0809$+$483}
\tablewidth{0pt}
\tablehead{\colhead{Features} & \colhead{$\lambda_{\rm rest}$} &
\colhead{$\lambda_{\rm obs}$} &  
\colhead{Flux (${\rm erg}/{\rm sec}/{\rm cm}^2$)} & 
\colhead{${\rm EW}_{\rm rest}$ (\AA)} & \colhead{$L$ ($h^{-2}\,{\rm erg}/{\rm sec}$)}}
\startdata
$[{\rm O\,II}]$ \dotfill & 3728.23 & 5361.29 & $(8.5 \pm 1.0)\times 10^{-17}$ &
$17.2 \pm 2.9$ & $6.8 \times 10^{39}$ \nl
H\,$\beta$  \dotfill & 4862.70 & 6982.72 & $(1.25 \pm 0.08)\times 10^{-16}$ & 
$9.9 \pm 0.7$  & $1.0 \times 10^{40}$ \nl
$[{\rm O\,III}]$ \dotfill & 4960.29 & 7129.92 & $< 4.7\times 10^{-18}$ &
$< 0.4 $ & $< 3.8 \times 10^{38}$ \nl
$[{\rm O\,III}]$ \dotfill & 5008.24 & 7198.84 & $< 4.7\times 10^{-18}$ &
$< 0.4 $ & $< 3.8 \times 10^{38}$ 
\enddata
\end{deluxetable}

\newpage

\begin{deluxetable}{p{0.6in}lrrrr}
\tablecaption{Spectral features identified in the DLA galaxy at $z=0.52530$
toward AO\,0235$+$164}
\tablewidth{0pt}
\tablehead{\colhead{Features} & \colhead{$\lambda_{\rm rest}$} &
\colhead{$\lambda_{\rm obs}$} &  
\colhead{Flux (${\rm erg}/{\rm sec}/{\rm cm}^2$)} & 
\colhead{${\rm EW}_{\rm rest}$ (\AA)} & \colhead{$L$ ($h^{-2}\,{\rm erg}/{\rm sec}$)}}
\startdata
$[{\rm Ne\,V}]$ \dotfill & 3426.48 & 5226.83 & $(5.5 \pm 0.2)\times 10^{-17}$ &
$2.7 \pm 0.5$ & $5.4 \times 10^{39}$ \nl
$[{\rm O\,II}]$ \dotfill & 3728.23 & 5685.65 & $(2.13 \pm 0.06)\times 10^{-16}$ &
$9.8 \pm 1.0$ & $2.1 \times 10^{40}$ \nl
H\,$\gamma$  \dotfill & 4341.69 & 6622.91 & $(6.0 \pm 0.3)\times 10^{-17}$ & 
$2.4 \pm 0.5$ & $5.9 \times 10^{39}$ \nl
$[{\rm O\,III}]$ \dotfill & 4364.44 & 6657.61 & $(2.7 \pm 0.2)\times 10^{-17}$ &
$1.1 \pm 0.3$ & $2.6 \times 10^{39}$ \nl
H\,$\beta$  \dotfill & 4862.70 & 7417.66 & $(9.2 \pm 0.5)\times 10^{-17}$ & 
$3.2 \pm 0.6$ & $9.0 \times 10^{39}$ \nl
$[{\rm O\,III}]$ \dotfill & 4960.29 & 7566.52 & $(1.11 \pm 0.07)\times 10^{-16}$ &
$3.8 \pm 0.7$ & $1.1 \times 10^{40}$ \nl
$[{\rm O\,III}]$ \dotfill & 5008.24 & 7639.67 & $(3.2 \pm 0.1)\times 10^{-16}$ &
$11.1 \pm 1.3$ & $3.1 \times 10^{40}$ 
\enddata
\end{deluxetable}

\begin{deluxetable}{p{0.6in}lrrrr}
\tablecaption{Spectral features identified in the DLA galaxy at $z=0.52631$
toward B2\,0827$+$243}
\tablewidth{0pt}
\tablehead{\colhead{Features} & \colhead{$\lambda_{\rm rest}$} &
\colhead{$\lambda_{\rm obs}$} &  
\colhead{Flux (${\rm erg}/{\rm sec}/{\rm cm}^2$)} & 
\colhead{${\rm EW}_{\rm rest}$ (\AA)} & \colhead{$L$ ($h^{-2}\,{\rm erg}/{\rm sec}$)}}
\startdata
$[{\rm O\,II}]$ \dotfill & 3728.23 & 5689.38 & $(7.9 \pm 0.7)\times 10^{-17}$ &
$39.6 \pm 3.5$ & $7.7 \times 10^{39}$ \nl
H\,$\beta$  \dotfill & 4862.70 & 7422.52 & $(2.6 \pm 0.4)\times 10^{-17}$ & 
$5.9 \pm 1.6$ & $2.5 \times 10^{39}$ \nl
$[{\rm O\,III}]$ \dotfill & 4960.29 & 7571.48 & $< 3.5 \times 10^{-18}$ &
$< 0.7$ & $< 3.4 \times 10^{38}$ \nl
$[{\rm O\,III}]$ \dotfill & 5008.24 & 7644.67 & $(1.7 \pm 0.1)\times 10^{-16}$ &
$30.8 \pm 4.4$  & $< 1.0 \times 10^{39}$\tablenotemark{a}
\tablenotetext{a}{The [O\,III] $\lambda\,5007$ feature coincides with the 
telluric A-band absorption.  The observed emission-line flux is subject to a
large systematic uncertainty after the correction for the atmosphere 
absorption.  The upper limit to the intrinsic line luminosity was estimated 
based on the measured [O\,III] $\lambda\,4959$ line luminosity, which is not 
contaminated by telluric absorption, and the expected 1:3 flux ratio between 
the two forbidden transitions.} 
\enddata
\end{deluxetable}

\newpage

\begin{deluxetable}{p{0.6in}lrrrr}
\tablecaption{Spectral features identified in the DLA galaxy at $z=0.6120$
toward LBQS\,0058$+$019}
\tablewidth{0pt}
\tablehead{\colhead{Features} & \colhead{$\lambda_{\rm rest}$} &
\colhead{$\lambda_{\rm obs}$} &  
\colhead{Flux (${\rm erg}/{\rm sec}/{\rm cm}^2$)} & 
\colhead{${\rm EW}_{\rm rest}$ (\AA)} & \colhead{$L$ ($h^{-2}\,{\rm erg}/{\rm 
sec}$)}}
\startdata
$[{\rm O\,II}]$ \dotfill & 3728.23 & 6010.01 & $(2.5\pm 0.2)\times 10^{-17}$ & 
$2.6 \pm 1.1$  & $2.8 \times 10^{39}$
\enddata
\end{deluxetable}

\begin{deluxetable}{p{1.05in}crcrrrlrcrr}
\tabletypesize{\scriptsize}
\rotate
\tablecaption{Summary of known physical properties of $z<1$ DLA systems}
\tablewidth{0pt}
\tablehead{\colhead{} & \multicolumn{6}{c}{Galaxies} & \colhead{} 
& \multicolumn{4}{c}{Absorbers} \\
\cline{2-7} 
\cline{9-12} \\
\colhead{} & \colhead{} & \colhead{$\rho$} & 
\colhead{$M_{AB}(B)$} & \colhead{$V_{\rm term}$} & 
\colhead{$M_{\rm dyn}$\tablenotemark{a}} & \colhead{} &
\colhead{} &  
\colhead{} & \colhead{$\log N(\hI)$} & \colhead{$\log N({\rm Fe\,II})$} & 
\colhead{} \\
\colhead{Sightline} & \colhead{$z_{\rm gal}$} & \colhead{($h^{-1}$\,kpc)} & 
\colhead{$-5\,\log\,h$} & \colhead{(${\rm km}/{\rm s}$)} & 
\colhead{$(10^{11}\,h^{-1}\,{\rm M}_\odot)$} &
\colhead{$[{\rm O}/{\rm H}]$\tablenotemark{b,c}} &
\colhead{} & 
\colhead{$z_{\rm DLA}$} & \colhead{(cm$^{-2}$)} & 
\colhead{(cm$^{-2}$)} & \colhead{$Z$} \\
\colhead{(1)} & \colhead{(2)} & \colhead{(3)} & \colhead{(4)} & \colhead{(5)}
& \colhead{(6)} & \colhead{(7)} & & \colhead{(8)} &  \colhead{(9)} & 
\colhead{(10)} & \colhead{(11)}}
\startdata
PKS0439$-$433 \dotfill  & 0.1010 &  5.3 & $-19.6$ & $233\pm 55$ & 
                        $1.2\pm 0.6$ & $0.45\pm 0.15$ & &
                       0.1009 & $19.85\pm 0.10$ & $14.58\pm 0.06$ & 
                       $-0.20\pm 0.30$ \nl
Q0738$+$313 \dotfill    & 0.2222 & 13.5 & $-17.7$ &        ...  & 
                                ...   & ... & &
                       0.2212 & $20.90\pm 0.08$ &  ... & ... \nl
Q0809$+$483 \dotfill    & 0.4375 &  5.9 & $-20.3$ & $320\pm 97$ &  
                        $1.8\pm 1.1$ & $\ge 0.36$ & & 
                       0.4368 & $20.80\pm 0.20$ &  ... & ... \nl
AO\,0235$+$164 \dotfill & 0.5253 &  9.2 & $-20.3$ &        ...  & 
                       ... & $-0.24\pm 0.15$ & &
                       0.5243 & $21.70\pm 0.09$ & $15.3\pm 0.4$ & 
                       $-0.60\pm 0.41$ \nl
B2\,0827$+$243 \dotfill & 0.5263 & 26.7 & $-20.0$ & $191\pm 33$ &
                        $2.3\pm 0.8$ & $\ge 0.06$ & &
                       0.5250 & $20.30\pm 0.10$ & $14.74\pm 0.04$ & 
                       $-0.49\pm 0.30$ \nl
LBQS0058$+$0155 \dotfill& 0.6120 &  5.5 & $-17.6$ &        ...  & 
                                  ...   & ... & &
                       0.6118 & $20.08\pm 0.18$ & $15.24\pm 0.10$ & 
                       $-0.02\pm 0.20$ \nl
\hline
HE1122$-$1649 \dotfill  & 0.69\tablenotemark{d} & 17.7 & $-18.8$ & ... & ... & ... & &
                       0.6819 & $20.45\pm 0.15$ & $14.56\pm 0.05$ & 
                       $-1.19\pm 0.15$ \nl
EX0302$-$2223 \dotfill  & 1.00\tablenotemark{d} & 18.5 & $-19.3$ & ... & ... & ... & &
                       1.0095 & $20.36\pm 0.09$ & $14.67\pm 0.10$ & 
                       $-0.80\pm 0.14$ 
\enddata
\tablenotetext{a}{Enclosed dynamic mass evaluated at the location of the 
DLA.}
\tablenotetext{b}{$[{\rm O}/{\rm H}]=\log ({\rm O}/{\rm H}) - \log ({\rm O}/{\rm H})_\odot$ and $12 + \log({\rm O}/{\rm H}) = 8.66$ (Allende-Prieto, Lambert, 
\& Asplund 2001; Asplund \etal\ 2004).}
\tablenotetext{c}{The oxygen abundances are derived using the $R_{23}$ index
and may be systematically higher by 0.2--0.5 dex.}
\tablenotetext{d}{Measurement based on photometric redshift techniques (Chen 
\& Lanzetta 2003).}
\end{deluxetable}

\clearpage

\begin{figure}
\plotone{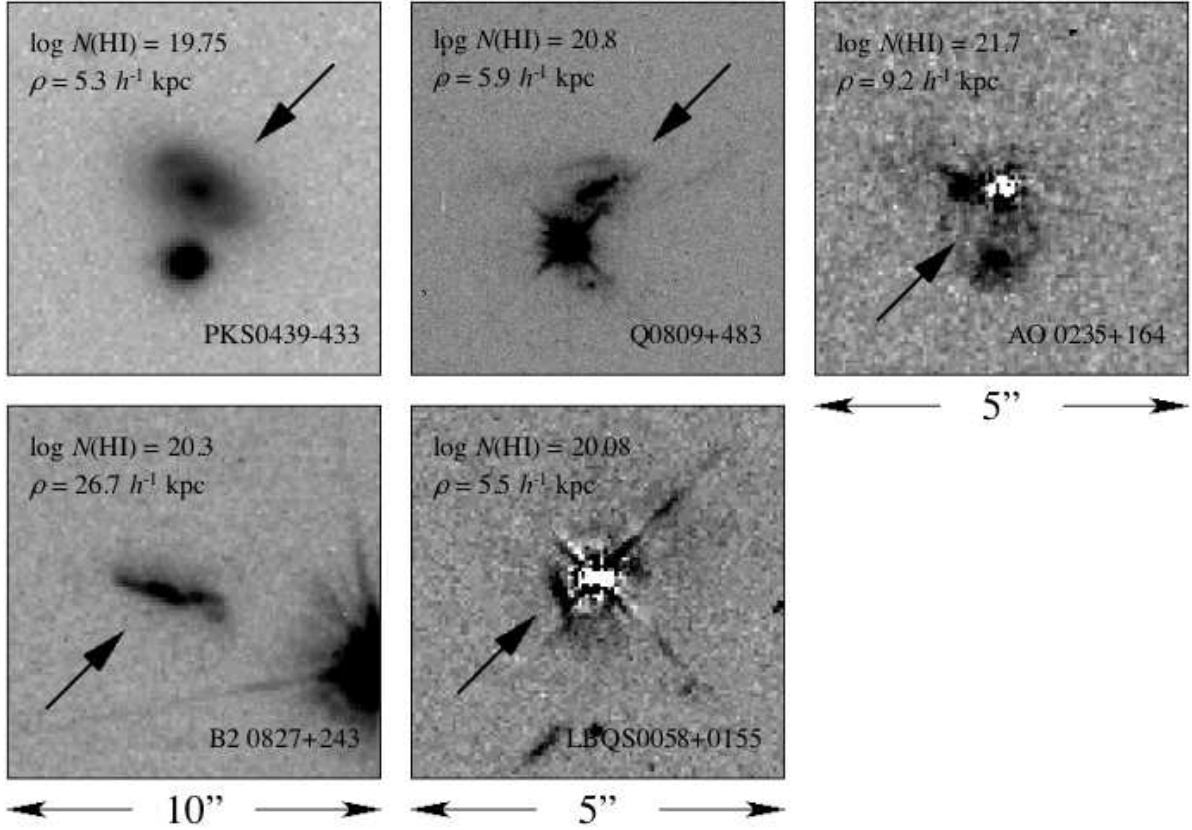}
\caption[]{Direct images of five DLA galaxies at $z<0.65$, showing a range of
morphologies.  The images are 10 arcsec on a side for the fields toward 
PKS0439$-$433 and B2\,0827$+$243 and 5 arcsec for the rest.  Field orientation 
is arbitrary.  All but the image toward PKS0439$-$433 were obtained with 
HST/WFPC2 using the F702W filter.  The image toward PKS0439$-$433 was obtained 
using the Tek\#5 CCD camera on the du Pont telescope at Las Campanas.  The 
light from the background QSOs toward AO\,0235$+$164 and LBQS0058$+$019 have 
been subtracted to bring out the faint features of the absorbing galaxies.}
\end{figure}

\clearpage

\begin{figure}
\plotone{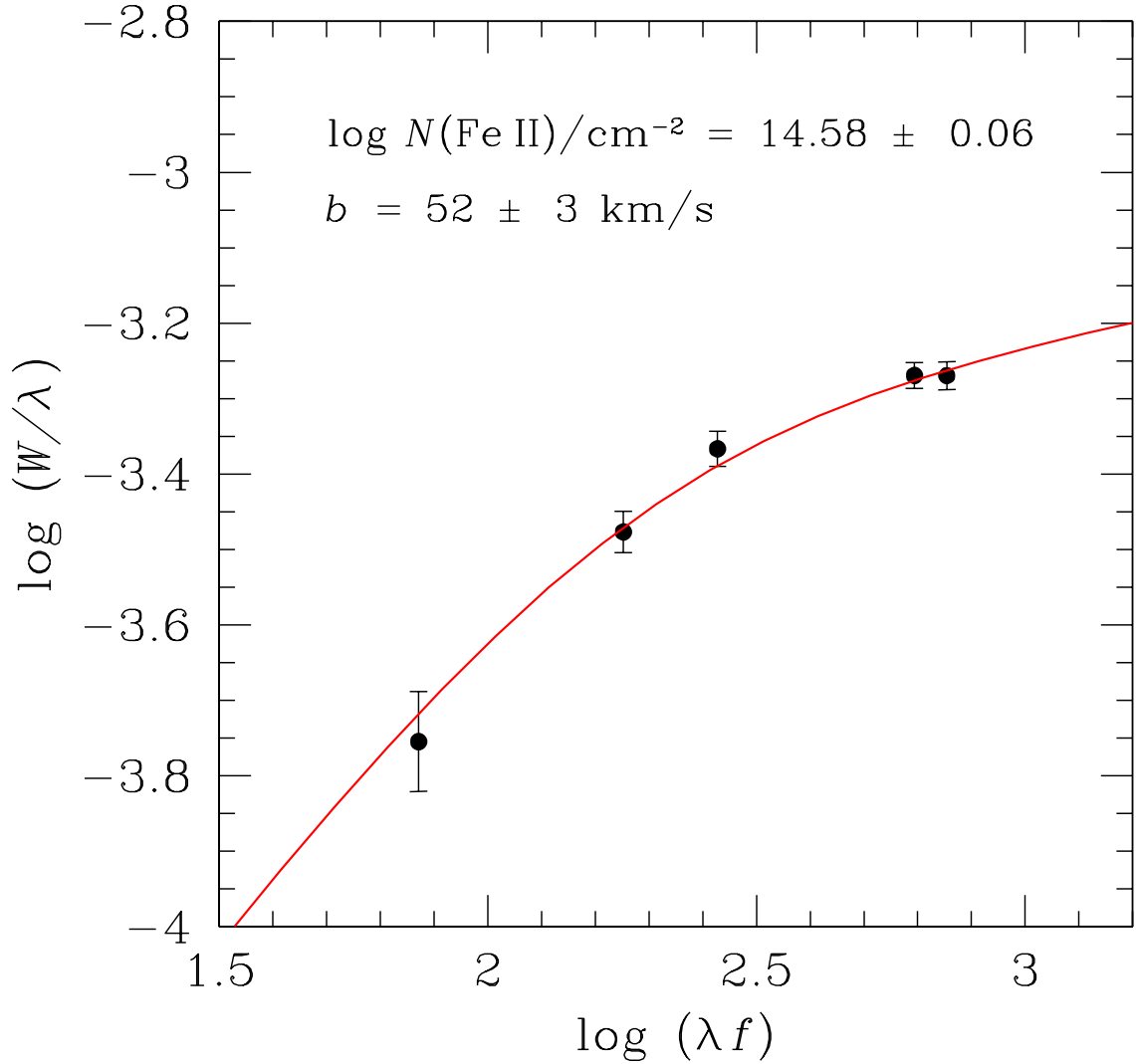}
\caption[]{Curve-of-growth analysis of the Fe\,II series for the DLA at 
$z=0.101$.  Solid points are measured rest-frame absorption equivalent widths
of the five Fe\,II transitions presented in Figure 3.  The solid curve 
represents the best-fit model with $\log N({\rm Fe\,II})=14.58\pm 0.06$ and a 
velocity width of $b=52\pm 3\,{\rm km} {\rm s}^{-1}$.}
\end{figure}

\clearpage

\begin{figure}
\plotone{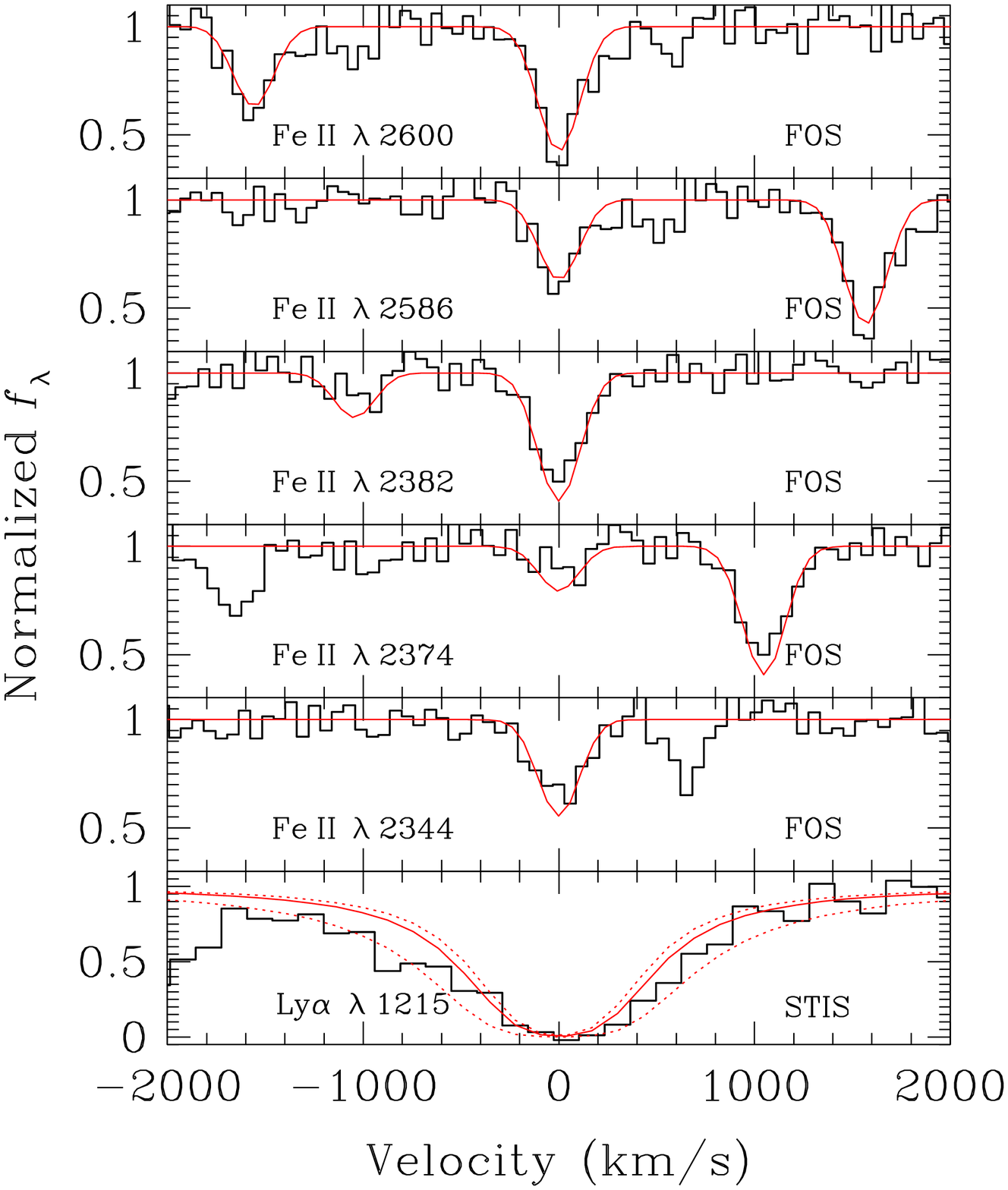}
\caption[]{Observed absorption-line profiles of five Fe\,II transitions and the
hydrogen \lya\ transition (histograms) of the DLA toward PKS0439$-$433.  The 
best-fit Voigt models with $N({\rm Fe\,II})=14.58\pm 0.06$ from a 
curve-of-growth analysis and $N(\hI)=19.85\pm 0.10$ from VPFIT are presented in
thin curves for comparison.  The 1-$\sigma$ uncertainty in $N(\hI)$ is also
presented in thin dotted curves at the bottom panel.  Zero velocity corresponds
to the absorber redshift $z_{\rm DLA}=0.10088$.}
\end{figure}

\clearpage

\begin{figure}
\plotone{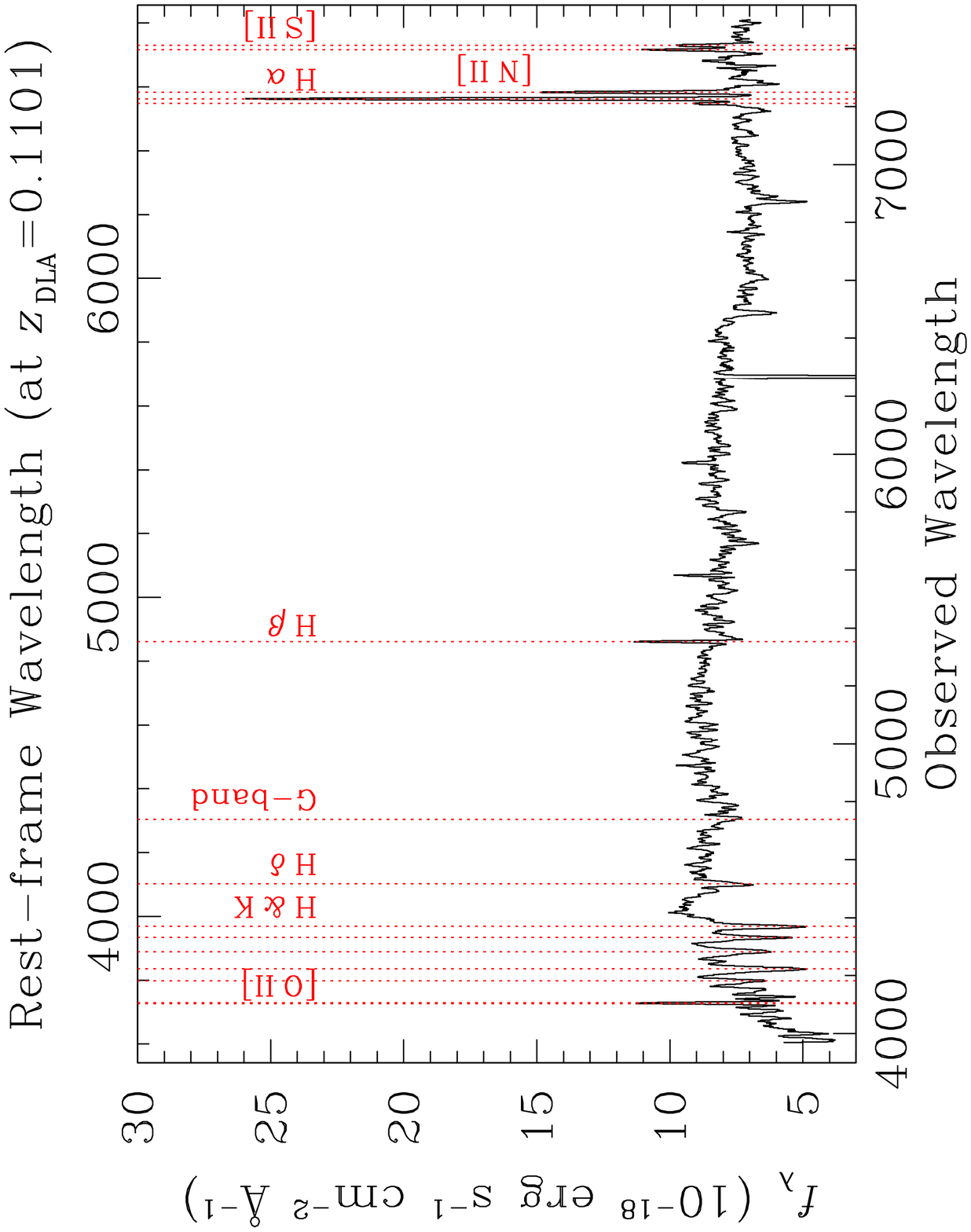}
\caption[]{The summed, extracted spectrum of the DLA galaxy at $z=0.10104$ 
toward PKS0439$-$433, covering a rest-frame wavelength range from 3550 \AA\ 
through 6860 \AA.  No extinction correction has been applied to the data.  The 
spectrum exhibits both prominent absorption- and emission-line features (dotted
lines), indicating a moderately young stellar population and metal-enriched 
ISM.}
\end{figure}

\clearpage

\begin{figure}
\plotone{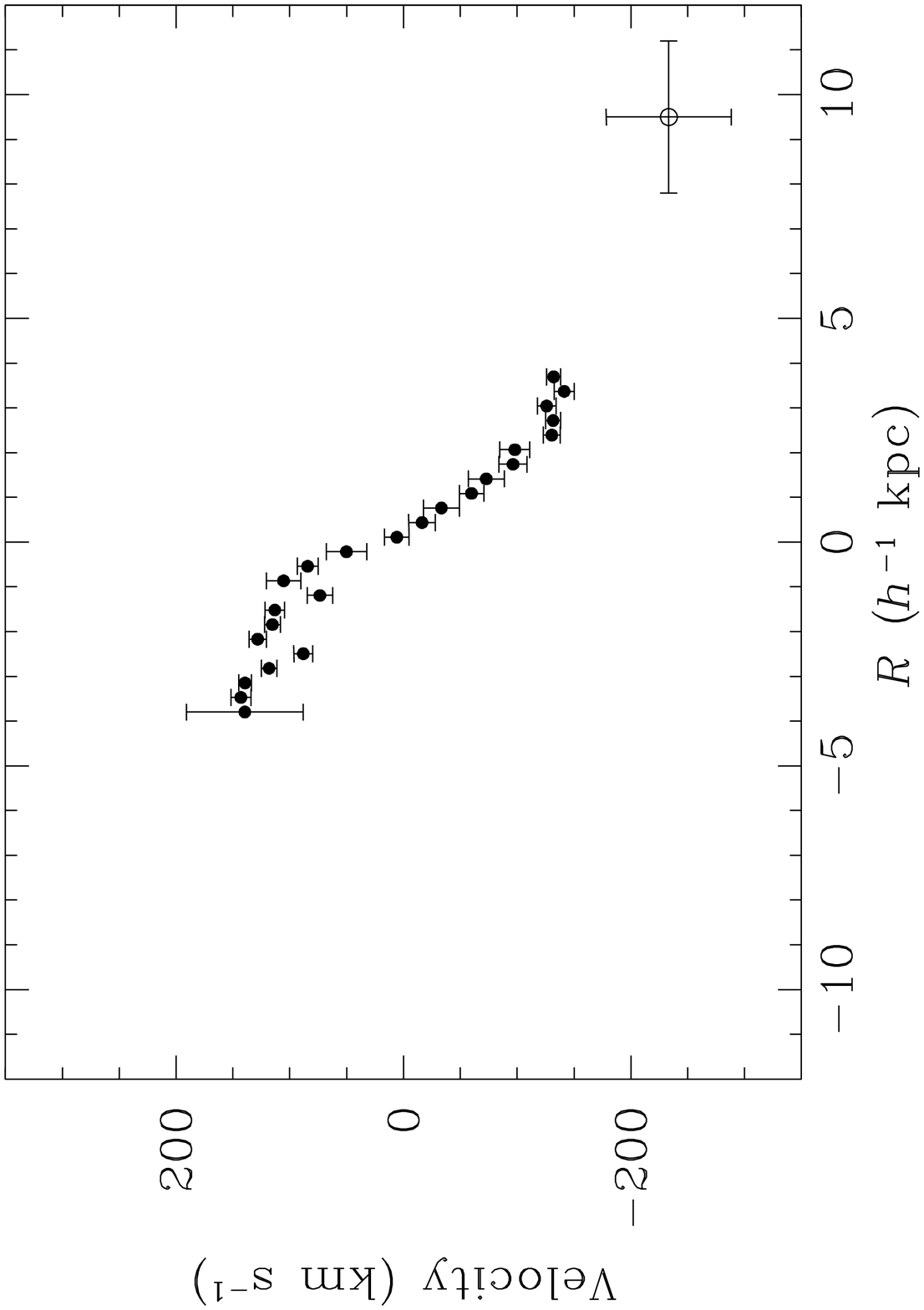}
\caption[]{Rotation velocity measurements of the DLA galaxy at $z=0.1010$ 
toward PKS0439$-$433 versus galactocentric radius $R$ along the disk (solid 
points).  The velocity measurements presented in the plot has been corrected 
for the inclination of the stellar disk.  The open circle shows the relative 
motion of the absorber with respect to the systematic velocity of the absorbing
galaxy, which has also been deprojected to the stellar disk.}
\end{figure}

\clearpage

\begin{figure}
\plotone{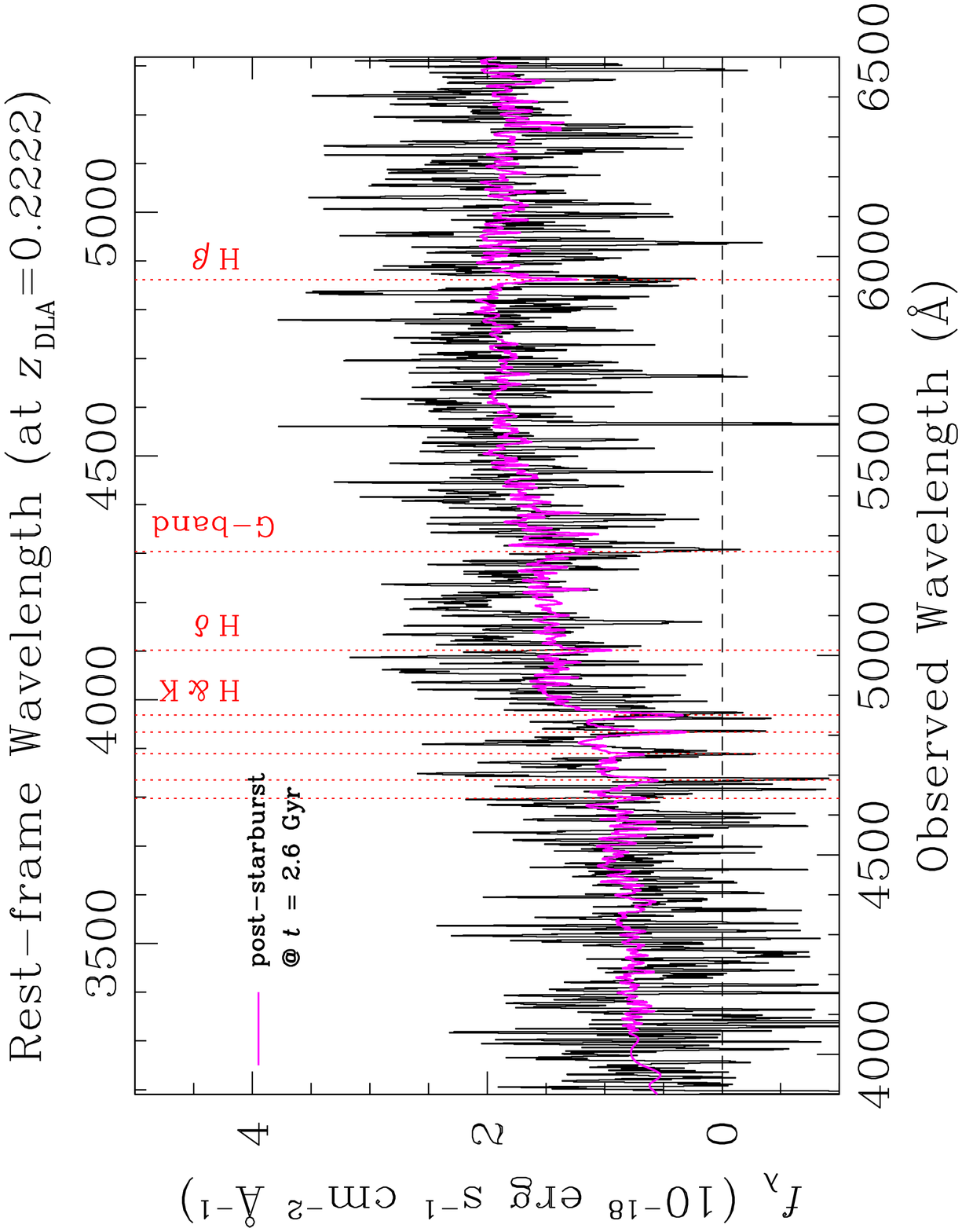}
\caption[]{The summed, extracted spectrum of the DLA galaxy at $z=0.2222$ 
toward Q0738$+$313 (OI\,363), covering rest-frame wavelengths from 3200 \AA\ 
through 5300 \AA.  No emission-line features, but only numerous absorption 
lines are detected in the spectrum.  For comparison, we have included a 2.6-Gyr
old model spectrum (the magenta curve) generated by the Bruzual \& Charlot 
stellar population synthesis code (2003) for an exponentially declining SFR of 
e-folding time $\tau = 0.3$ Gyr and solar metallicity that best describes the 
spectrum of the galaxy.}
\end{figure}

\clearpage

\begin{figure}
\plotone{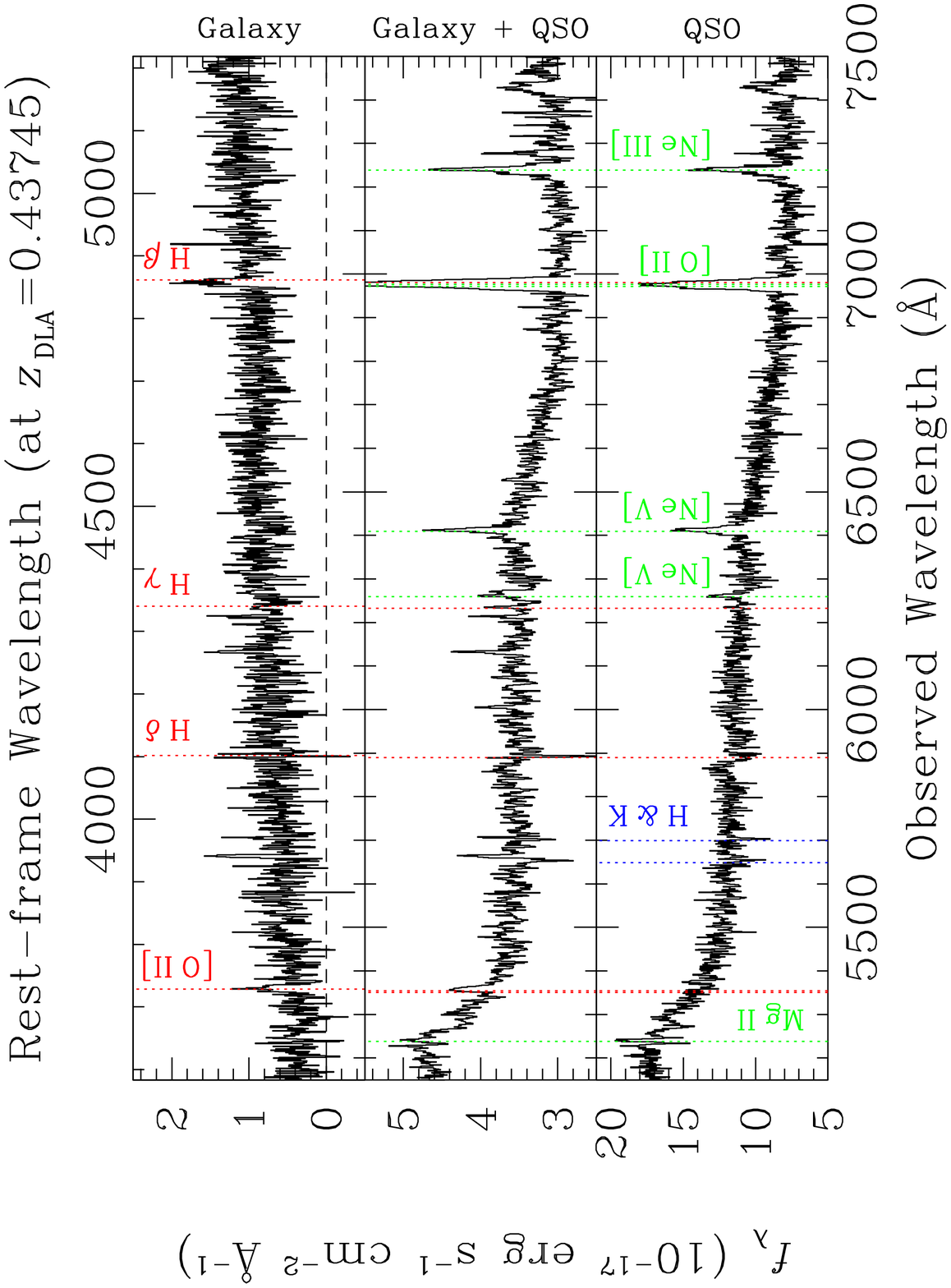}
\caption[]{The summed, extracted spectrum of the DLA galaxy at $z=0.43745$ 
toward Q0809$+$483, covering a rest-frame wavelength range from 3600 \AA\ 
through 5200 \AA.  No extinction correction has been applied to the data.  The 
panels show from bottom to top the spectrum of the background QSO, the 
extracted galaxy spectrum with contaminating QSO light due to their close 
proximity on the sky, and the difference spectrum after subtracting off 
contaminating QSO features from the extracted galaxy spectrum.  The residual 
fluxes above the continua at the locations of the [Ne\,V] and [Ne\,III] (dashed
lines) transitions from the QSO in the difference spectrum are consistent with 
zero line fluxes, lending much confidence in the detection of H$\beta$ in the 
galaxy.  The prominent emission-line features of [O\,II] and H$\beta$ (dotted
lines) and a lack of strong absorption features together suggest a moderately 
young stellar population and metal-enriched ISM in the galaxy.  The QSO 
spectrum shown in the bottom panel also exhibits Ca\,II H\&K absorption 
(dash-dotted lines) at 21 \kms\ blueshifted from the systematic redshift of the
galaxy.}
\end{figure}

\clearpage

\begin{figure}
\plotone{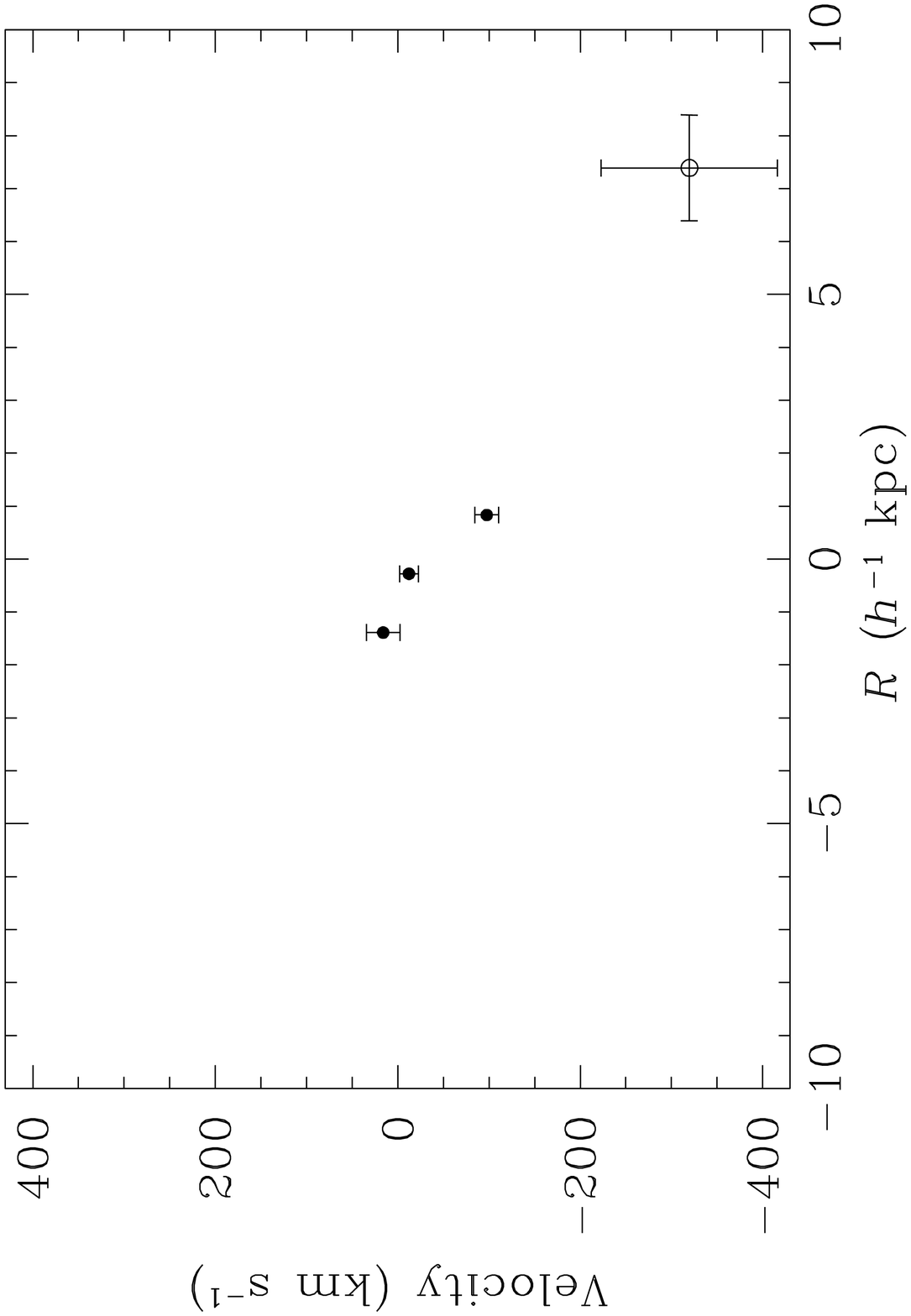}
\caption[]{Rotation velocity measurements of the DLA galaxy at $z=0.43745$ 
toward Q0809$+$483 versus galactocentric radius $R$ along the disk (solid 
points).  The velocity measurements presented in the plot has been corrected 
for the inclination of the stellar disk.  The open circle shows the relative 
motion of the absorber with respect to the systematic velocity of the absorbing
galaxy, which has also been deprojected to the stellar disk.}
\end{figure}

\clearpage

\begin{figure}
\plotone{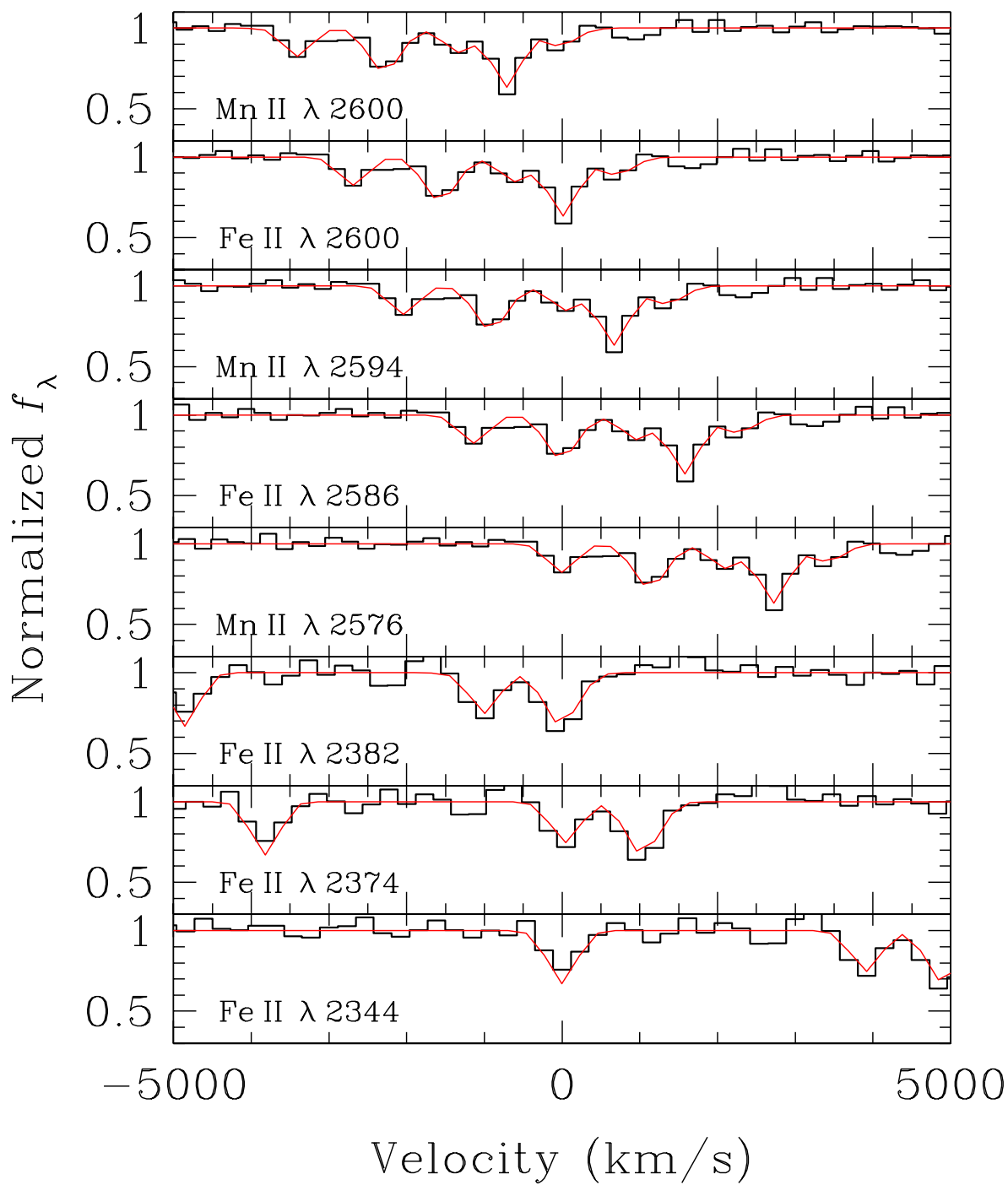}
\caption[]{Observed absorption-line profiles of three Mn\,II and five Fe\,II 
transitions (histograms) of the DLA toward AO\,0235$+$164.  The best-fit Voigt
models with $N({\rm Fe\,II})=15.3\pm 0.4$ and $N({\rm Mn\,II})=13.79\pm 0.08$ 
from VPFIT are presented in thin curves for comparison.  Zero velocity 
corresponds to the absorber redshift $z_{\rm DLA}=0.52429$.}
\end{figure}

\clearpage

\begin{figure}
\plotone{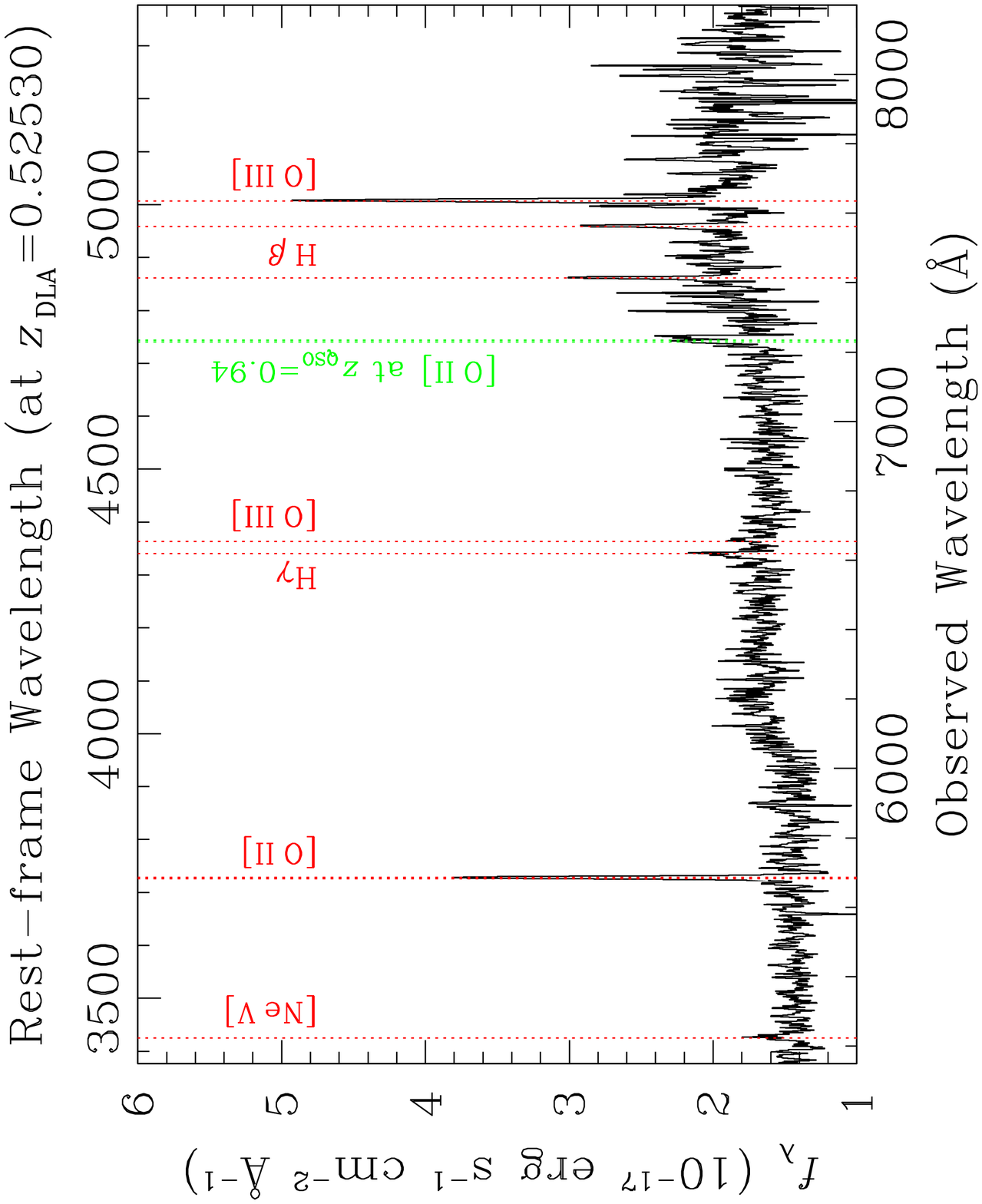}
\caption[]{The summed, extracted spectrum of the DLA galaxy at $z=0.52530$ 
toward AO\,0235$+$164, covering a rest-frame wavelength range from 3400 \AA\ 
through 5300 \AA.  No extinction correction has been applied to the data.  In
addition to the emission-line features commonly observed in an H\,II region, 
the spectrum exhibits a prominent [Ne\,V] emission feature that is 
characteristic of a starburst galaxy, as well as strong 4000-\AA\ flux 
discontinuity that is characteristic of an evolved stellar population.  
Contaminating [O\,II] emission feature (dashed line) from the background QSO 
$z_{\rm em} = 0.94$ is also observed.}
\end{figure}

\clearpage

\begin{figure}
\plotone{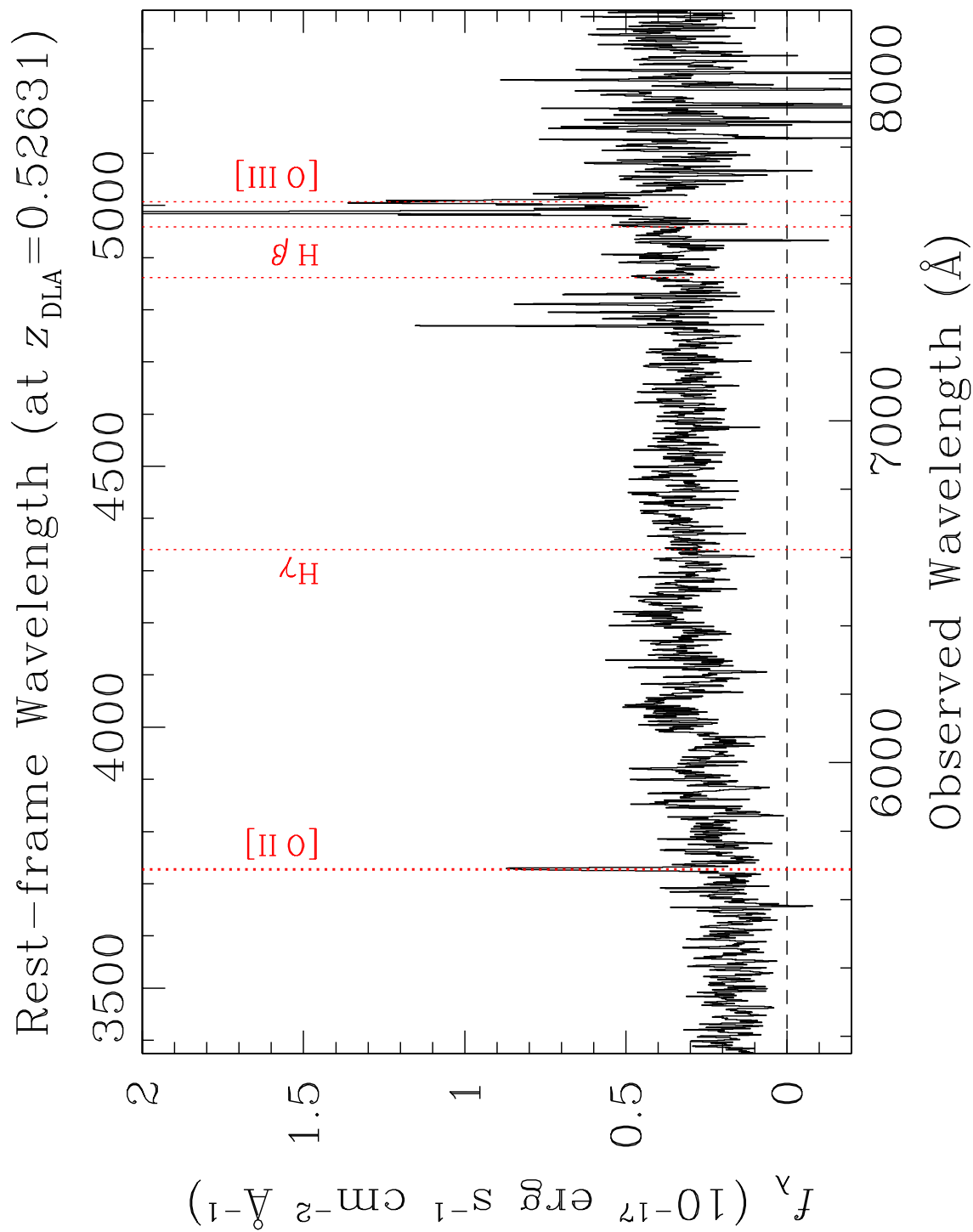}
\caption[]{The summed, extracted spectrum of the DLA galaxy at $z=0.52631$ 
toward B2\,0827$+$243, covering a rest-frame wavelength range from 3400 \AA\ 
through 5400 \AA.  No extinction correction has been applied to the data.}
\end{figure}

\clearpage

\begin{figure}
\plotone{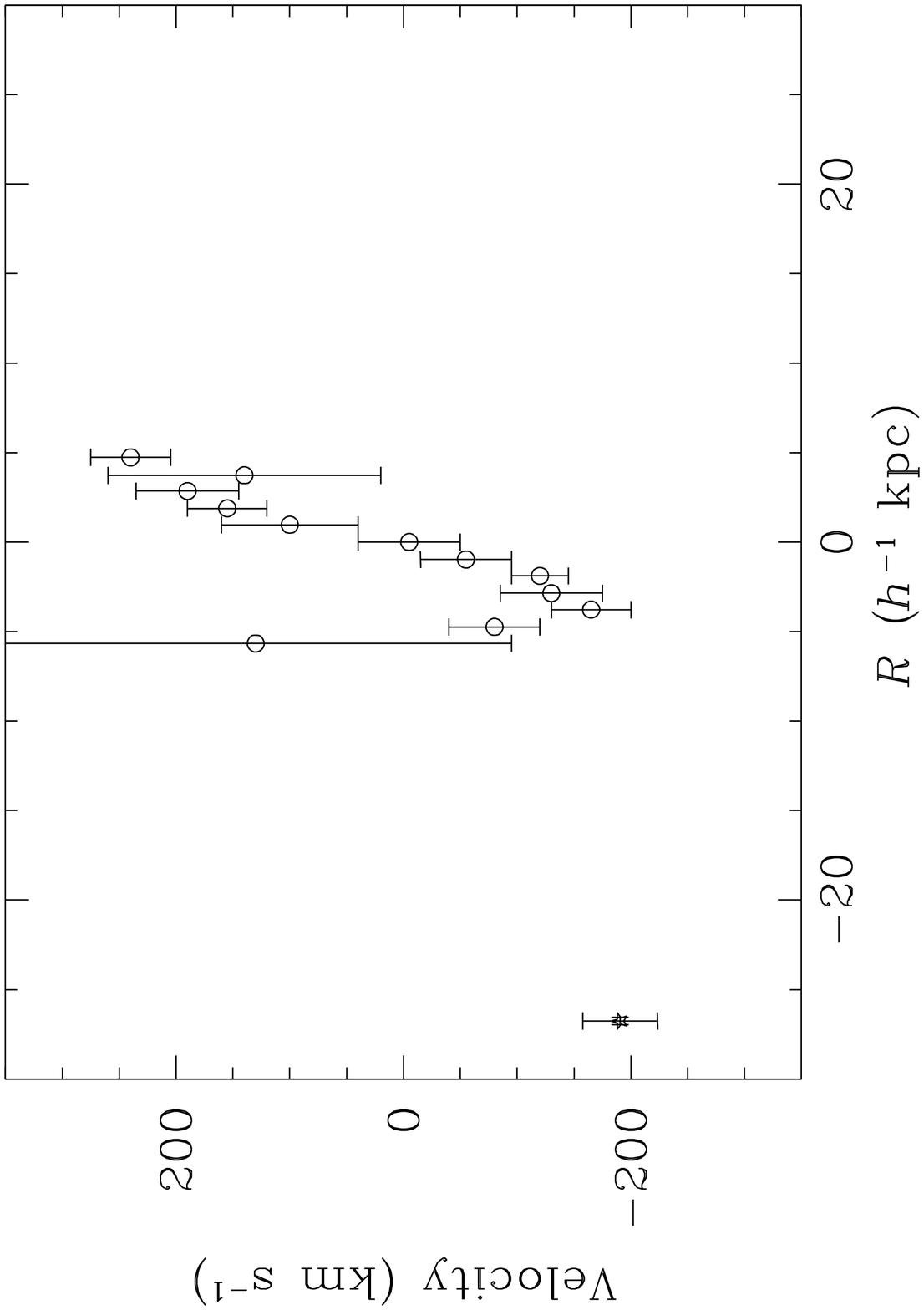}
\caption[]{Rotation velocity measurements of the edge-on DLA galaxy at $z =
0.52631$ toward B2\,0827$+$243 versus galactocentric radius $R$ along the disk 
(open points).  The velocity measurements presented in the plot are extracted 
from Steidel \etal\ (2002).  The open star shows the relative motion of the 
absorber with respect to the systematic velocity of the absorbing galaxy.  No 
correction has been made for the disk orientation because of its edge-on nature
and because the QSO sightline coincides with the extension of the edge-on disk 
(see Figure 1).}
\end{figure}

\clearpage

\begin{figure}
\plotone{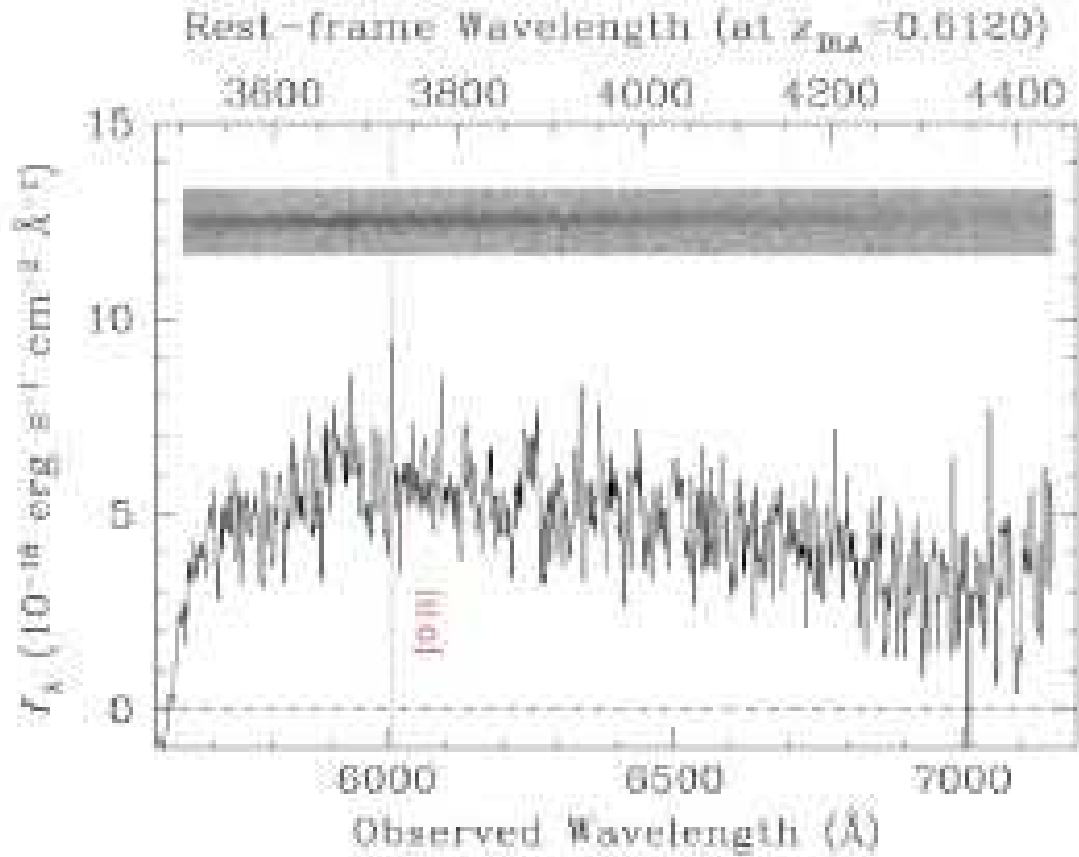}
\caption[]{The summed, extracted spectrum of the DLA galaxy at $z=0.6120$
toward LBQS0058$+$0155, covering a rest-frame wavelength range from 3500 \AA\ 
through 4450 \AA.  No extinction correction has been applied to the data.  The
inset shows the two-dimensional spectral image of the galaxy that matches the
wavelength scale of the one-dimensional spectrum.  The [O\,II] emission feature
is evident in the two-dimensional image.}
\end{figure}

\clearpage

\begin{figure}
\plotone{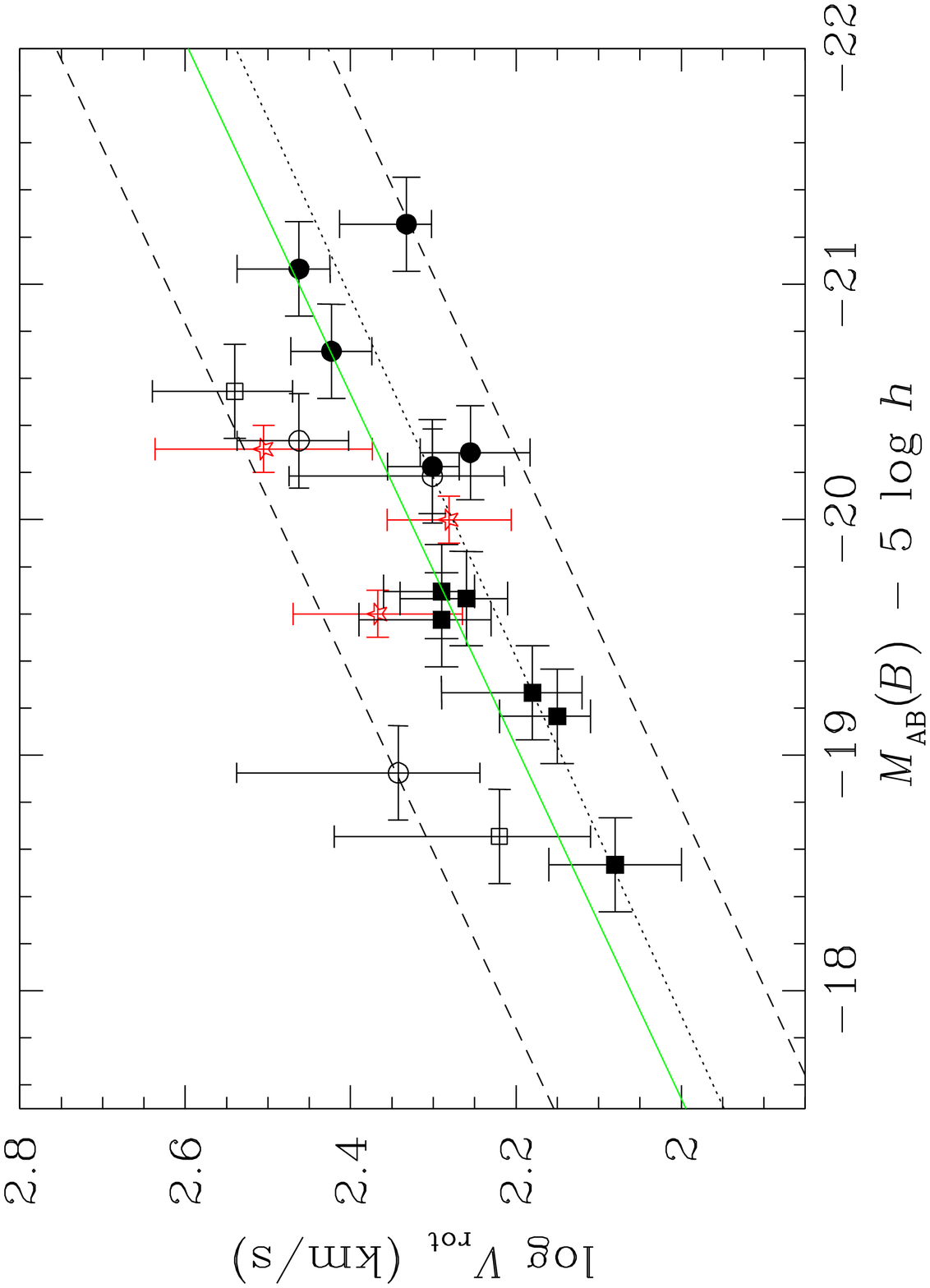}
\caption[]{Tully-Fisher relation of galaxies selected by $N(\hI)$ (open stars),
in comparison to field galaxies published in Vogt \etal\ (1997; solid points 
and squares).  The rotation velocity $V_{\rm rot}$ of each DLA system was 
measured from comparing the redshift of the DLA and the systematic redshift of 
the absorbing galaxy, corrected for the inclination of the disk and the 
relative orientation of the QSO sightline along the disk (see \S\ 3.1.3).  
Measurements for the field galaxies have been corrected for the $\Lambda$ 
cosmology and the symbols follow the presentation defined in Vogt et al.  The 
dotted line is the best-fit Tully-Fisher relation of the field sample from 
these authors.  The solid line represents the best-fit $B$-band relation from 
Pierce \& Tully (1988, 1992) based on \hI\ 21\,cm velocity measurements.  The 
corresponding 3-$\sigma$ confidence interval is indicated by the dashed lines.}
\end{figure}

\clearpage

\begin{figure}
\plotone{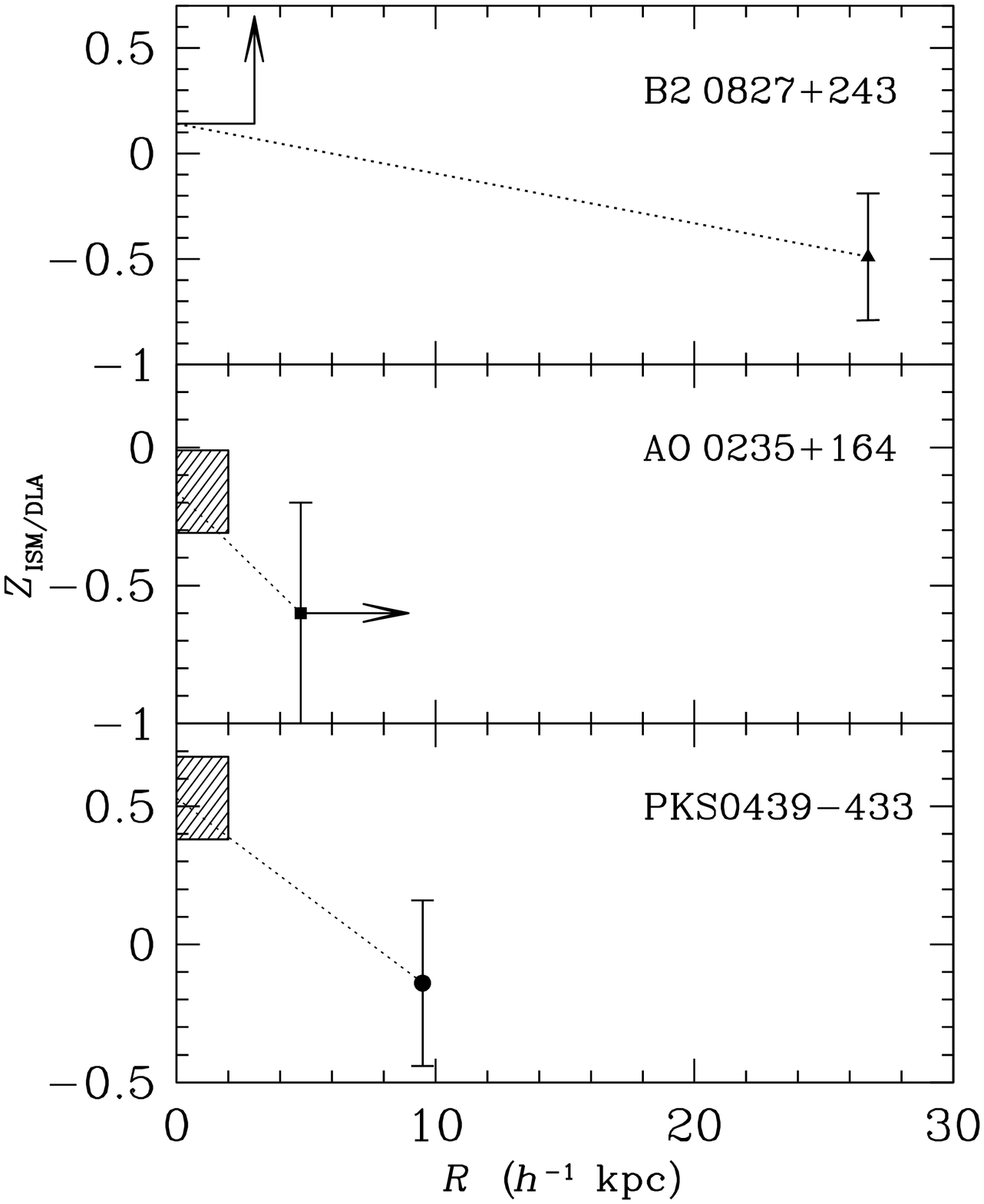}
\caption[]{Abundance decrement observed from the inner ISM of the absorbing 
galaxies to the neutral gas at large galactocentric radii for three DLA 
systems.  The shaded boxes indicate the mean oxygen abundance averaged over the
inner stellar disk, including uncertainties.  Only a lower limit to the oxygen 
abundance was available for the DLA galaxy toward B2\,0827$+$243.  The DLA 
toward AO\,0235$+$164 is presented at its projected distance to the absorbing 
galaxy, because the galaxy exhibits a complex morphology that cannot be 
represented by a simple disk model.  Metallicities of the ISM are oxygen
abundances derived using the $R_{23}$ index.  Including the uncertainties in
the calibration of the $R_{23}$ index may lower the decrement by 0.2--0.5 dex.
Metallicities of the absorbers are estimated based on the observed Fe 
abundance, corrected for dust depletion.  All abundace measurements are 
normalized to their respective solar values.}
\end{figure}

\clearpage

\begin{figure}
\plotone{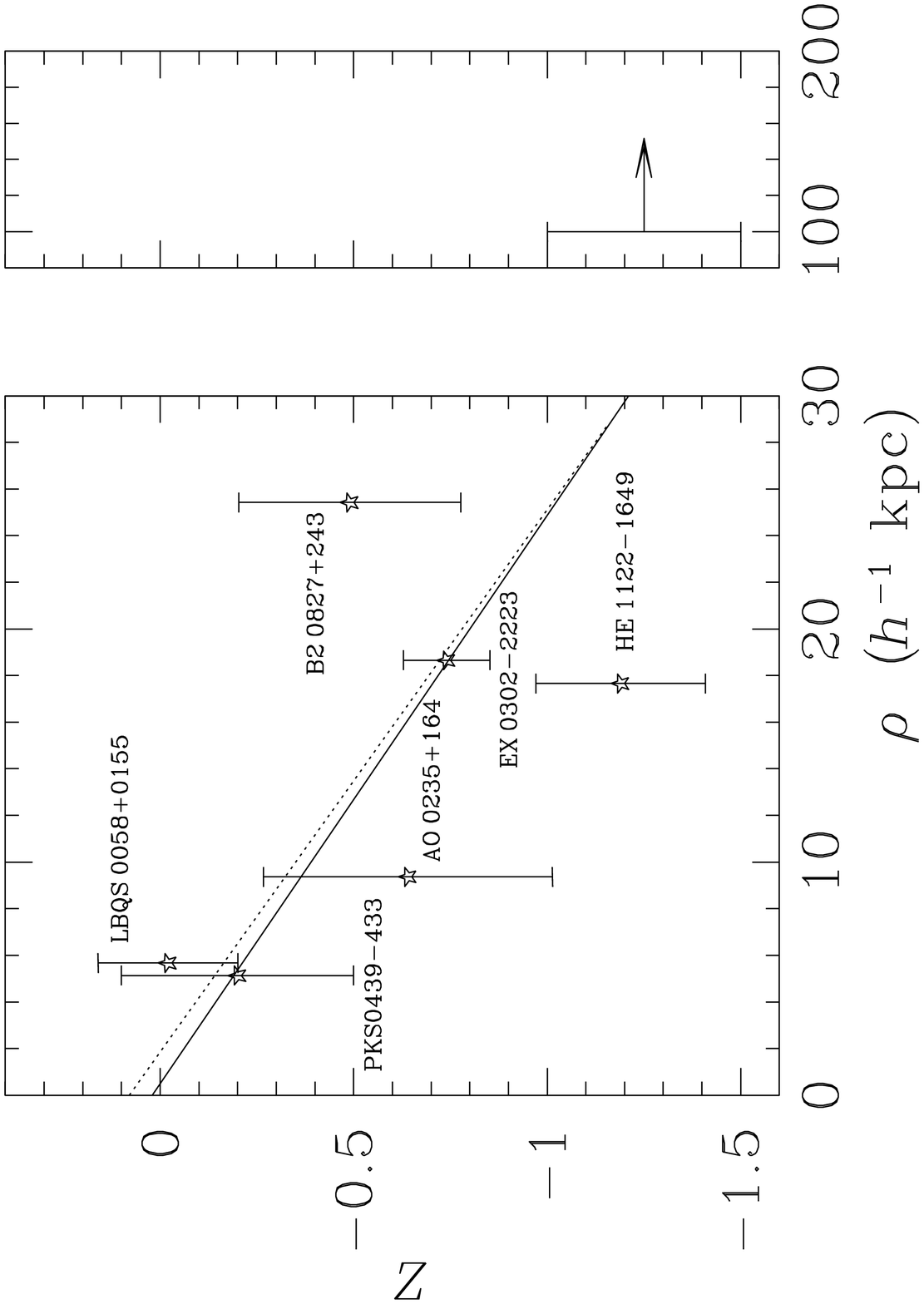}
\caption[]{Metallicity of neutral gas $Z$ versus galaxy impact parameter 
$\rho$.  Open stars are derived metal abundances after corrections for dust 
depletion with corresponding 1-$\sigma$ measurement errors.  Solid line 
represents the best-fit correlation between $Z$ and $\rho$ with a gradient 
$-0.041\pm 0.012$ dex per kiloparsec, in comparison to the observed gradient of
the H\,II region oxygen abundance in M101 (dotted line; Kennicutt \etal\ 
2003).  The intergalactic medium metallicity estimated based on five $z<0.5$ 
\lya\ absorbers with $N(\hI) = 10^{14-16}$ \cmjj\ from Prochaska \etal\ (2004)
is placed at distance beyond $\rho=100\ h^{-1}$ kpc in the right panel.}
\end{figure}

\clearpage

\begin{figure}
\plotone{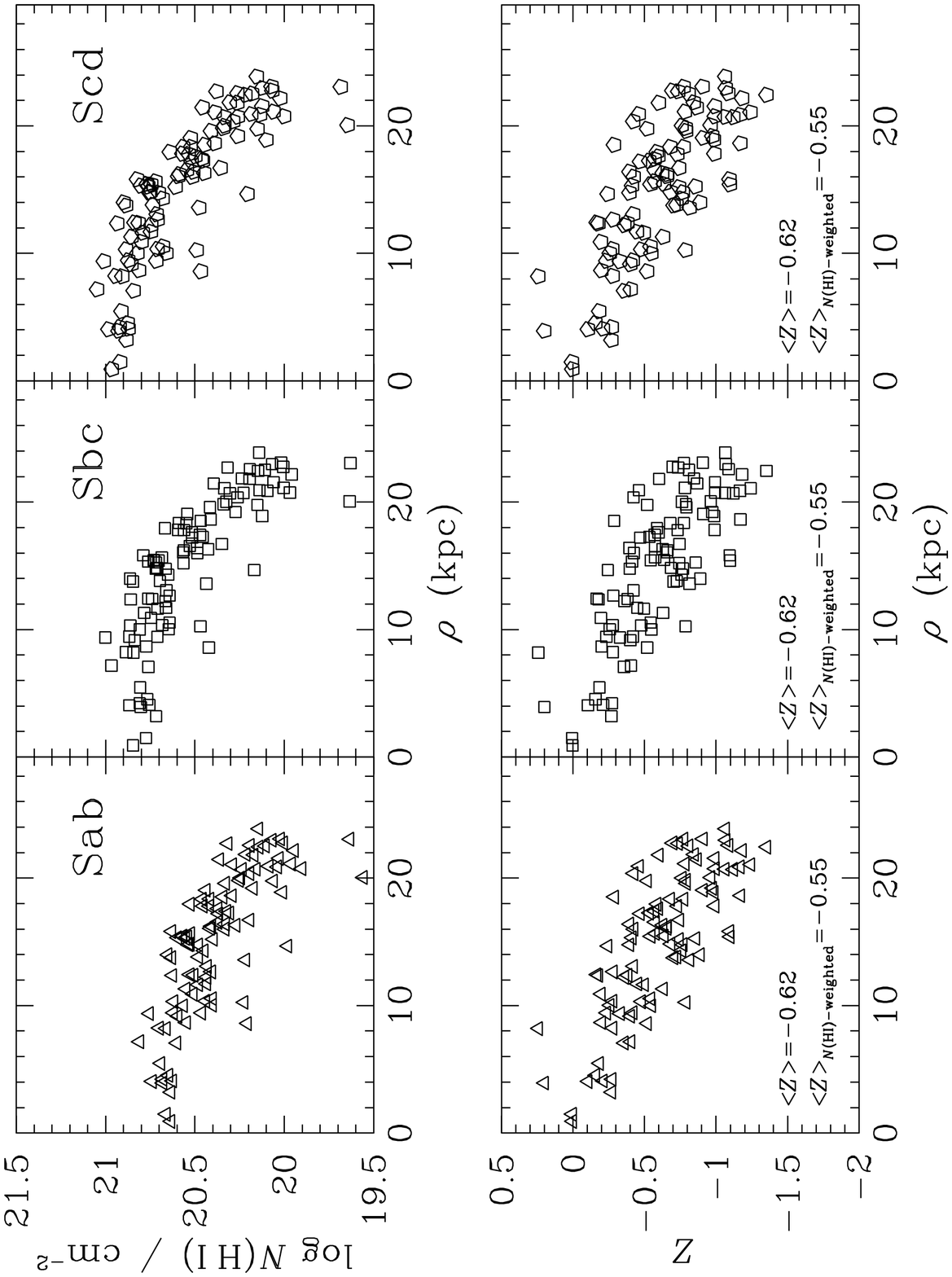}
\caption[]{Monte-Carlo simulation results of the $N(\hI)$ (top panels) and 
metallicity (bottom panels) distributions of a mocked sample of 100 DLA 
systems randomly drawn from the adopted mean profiles of local disk galaxies of
three different morphological types (indicated in the upper-right corner in the
top panels).  The predicted mean metallicity with or without weighted by 
$N(\hI)$ for each galaxy type is presented in the lower-left corner of the
bottom panels.}
\end{figure}

%


\newpage

\begin{references}

\reference{} Albert, C. E., Blades, J. C., Morton, D. C., Lockman, F. J., 
             Proulx, M., \& Ferrarese, L. 1993, ApJS, 88, 81

\reference{} Allende-Prieto, C., Lambert, D. L., \& Asplund, M. 2001, ApJ, 556,
             L63

\reference{} Asplund, M., Grevesse, N., Sauval, A. J., Allende Prieto, C., 
             \& Kiselman, D. 2004, A\&A, 417, 751

\reference{} Bahcall, J. N., et al.\ 1993, ApJS, 87, 1

\reference{} Baldwin, J.A., Phillips, M.M., \& Terlevich, R. 1981, PASP, 93, 5

\reference{} Bechtold, J., Dobrzycki, A., Wilden, B., Morita, M., Scott, J., 
             Dobrzycka, D., Tran, K., \& Aldcroft, T. L. 2002, ApJS, 140, 143

\reference{} Boiss\'e, P., Le Brun, V., Bergeron, J., \& Deharveng, J.-M. 1998,
             A\&A, 333, 841

\reference{} Bowen, D.V., Tripp, T.M., \& Jenkins, E.B. 2001, AJ, 121, 1456

\reference{} Bresolin, F., Garnett, D.R., \& Kennicutt, R.C. 2004,
             ApJ, in press (astro-ph/0407065)

\reference{} Broeils, A.H., \& van Woerden, H. 1994, A\&AS, 107, 129

\reference{} Brown, R.L. \& Mitchell, K.J. 1983, ApJ, 264, 87

\reference{} Bruzual, G. \& Charlot, S. 2003, MNRAS, 344, 1000

\reference{} Burbidge, E.M., Caldwell, R.D., Smith, H.E., Liebert, T., \& 
	     Spinrad, H. 1976, ApJ, 205, L117

\reference{} Burbidge, E.M., Beaver, E.A., Cohen, R.D., Junkkarinen, V.R.,
	     \& Lyons, R.W. 1996, AJ, 112, 2533

\reference{} Calzetti, D., Kinney, A.L., Storchi-Bergmann, T. 1994, ApJ, 429,
             582

\reference{} Cardelli, J.A., Clayton, G.C., \& Mathis, J.S. 1989, ApJ, 345, 245

\reference{} Cayatte, V., Kotanyi, C., Balkowski, C., \& van Gorkom, J.H.
             1994, AJ, 107, 1003

\reference{} Chen, H.-W., Lanzetta, K. M., Webb, J. K., \& Barcons, X. 2001, 
             ApJ, 559, 654

\reference{} Chen, H.-W. \& Lanzetta, K. M. 2003, ApJ, 597, 706

\reference{} Cohen, J.G. 2001, AJ, 121, 1275

\reference{} Cohen, R.D., Smith, H.E., Junkkarinen, V.T., \& Burbidge, E.M. 
             1987, ApJ, 318, 577

\reference{} Cohen, R.D., Burbidge, E.M., Junkkarinen, V.T., Lyons, R.W., 
             \& Madejski, G. 1999, AAS, 194, 71.01

\reference{} Colbert, J. W. \& Malkan, M. A. 2002, ApJ, 566, 51


\reference{} de Blok, W. J. G., McGaugh, S. S., \& van der Hulst, J. M. 1996, 
             MNRAS, 283, 18

\reference{} Deharveng, L., Pe\~{n}a, M., Caplan, J., \& Costero, R. 2000, 
             MNRAS, 311, 329

\reference{} Dessauges-Zavadsky, M., Prochaska, J., \& D'Odorico, S. 2002, 
             A\&A, 391, 801

\reference{} D'Odorico, V. \& Molaro, P. 2004, A\&A, 415, 879

\reference{} Esteban, C., Peimbert, M., Torres-Peimbert, S., \&
             Escalante, V. 1998, MNRAS, 295, 401

\reference{} Evans, I. N. \& Koratkar, A. P. 2004, ApJS, 150, 73

\reference{} Ferguson, A. M. N., Gallagher, J.S., \& Wyse, R.F.G. 1998, AJ,
             116, 673

\reference{} Ferland, G. J. 2001, PASP, 113, 41


\reference{} Jenkins, E. B. 1986, ApJ, 304, 739

\reference{} Jerjen, H., Binggeli, B., \& Barazza, F. D. 2004, AJ, 127, 771

\reference{} Junkkarinen, V. T., Cohen, R. D., Beaver, E. A., Burbidge, E. M.,
             Lyons, R. W., \& Madejski, G. 2004, ApJ in press 
             (astro-ph/0407281)

\reference{} Khare, P., Kulkarni, V.P., Lauroesch, J.T., York, D.G., Crotts, 
             A.P.S., \& Nakamura, O. 2004, ApJ in press (astro-ph/0408139)

\reference{} Kennicutt, R.C. Jr. \& Garnett, D.R. 1996, ApJ, 456, 504

\reference{} Kennicutt, R.C. Jr. 1998a, ARA\&A, 36, 189

\reference{} Kennicutt, R.C. Jr. 1998b, ApJ, 498, 541

\reference{} Kennicutt, R.C. Jr., Bresolin, F., \& Garnett, D.R.  2003, ApJ,
             591, 801


\reference{} Kobulnicky, H. A. \& Zaritsky, D. 1999, ApJ, 511, 118

\reference{} Kobulnicky, H. A., Kennicutt, R. C., Jr., \& Pizagno, J. L. 1999, 
             ApJ, 514, 544

\reference{} Kulkarni, V.P., Fall, S.M., Lauroesch, J. T., Khare, P., \& 
             Truran, J. W. 2004, ApJ in press (astro-ph/0409234)

\reference{} Lacy, M., Becker, R. H., Storrie-Lombardi, L. J.\etal\ 2003, AJ,
             126, 2230

\reference{} Lane, W., Smette, A., Briggs, F., Rao, S., Turnshek, D., \& 
	     Meylan, G. 1998, AJ, 116, 26

\reference{} Lane, W. 2000, PhD. Thesis, Univ. of Groningen

\reference{} Lanzetta, K. M., Wolfe, A. M., \& Turnshek, D. A. 1995, ApJ, 440,
             435

\reference{} Le Brun, F., Bergeron, J., Boiss\'e, P., \& Deharveng, J.M. 1997,
	     A\&A, 321, 733

\reference{} Ledoux, C., Bergeron, J., \& Petitjean, P. 2002, A\&A, 385, 802

\reference{} Lu, L., Sargent, W. L. W., Barlow, T. A., Churchill, C. W., \& 
             Vogt, S. 1996, ApJS, 107, 475

\reference{} McGaugh, S. S. 2001, ApJ, 380, 140

\reference{} McWilliam, A. 1997, ARA\&A, 35, 503

\reference{} Meyer, D. M., Jura, M., \& Cardelli, J. A. 1998, ApJ, 493, 222

\reference{} Miller, E.D., Knezek, P.M., \& Bregman, J.N. 1999, ApJ, 510,
             L95

\reference{} Moos, H. W., \etal\ 2002, ApJS, 140, 3

\reference{} Petitjean, P., Theodore, B., Smette, A., \& Lespine, Y. 1996,
             A\&A, 313, 25

\reference{} Pettini, M., Smith, L. J., Hunstead, R. W., \& King, D. L. 1994, 
             ApJ, 426, 79

\reference{} Pettini, M., King, D. L., Smith, L. J., \& Hunstead, R. W. 1997, 
             ApJ, 478, 536

\reference{} Pettini, M., Ellison, S.L., Steidel, C.C., \& Bowen, D.V. 1999,
             ApJ, 510, 576

\reference{} Pettini, M., Ellison, S.L., Steidel, C.C., Shapley, A.E., \&
             Bowen, D.V. 2000, ApJ, 532, 65

\reference{} Pettini, M. \& Pagel, B.E.J. 2004, MNRAS, 348, L59

\reference{} Pierce, M. J. \& Tully, R. B. 1988, ApJ, 330, 579 

\reference{} Pierce, M. J. \& Tully, R. B. 1992, ApJ, 387, 47 


\reference{} Prochaska, J. X., Howk, J. C., O'Meara, J. M., Tytler, D., Wolfe, 
             A. M., Kirkman, D., Lubin, D., \& Suzuki, N. 2002, ApJ, 571, 693

\reference{} Prochaska, J. X., et al. 2003a, ApJ, 595, L9

\reference{} Prochaska, J. X., Gawiser, E., Wolfe, A. M., Cooke, J., \& Gelino,
             D. 2003b, ApJS, 147, 227

\reference{} Prochascka, J. X., Howk, J. C., \& Wolfe, A. M. 2003c, Nature, 
             423, 57

\reference{} Prochaska, J.X., Chen, H.-W., Howk, J.C., Weiner, B.J., \& 
             Mulchaey, J.S. 2004, ApJ in press (astro-ph/0408294)

\reference{} Rao \& Turnshek 1998, ApJ, 500, L115

\reference{} Rao \& Turnshek 2000, ApJS, 130, 1

\reference{} Rao, S.M., Nestor, D.B., Turnshek, D.A., Lane, W., Monier, E.M., 
             \& Bergeron, J. 2003, ApJ, 595, 94

\reference{} Rauch, M., Sargent, W. L. W., Barlow, T. A., \&  Simcoe, R. A.
             2002, ApJ, 576, 45

\reference{} Rola, C.S., Terlevich, E., \& Terlevich, R.J. 1997, MNRAS, 289, 
             419

\reference{} Schlegel, D.J., Finkbeiner, D.P., \& Davis, M. 1998, ApJ, 500,
             525

\reference{} Schulte-Ladbeck, R.E., Rao, S.M., Drozdovsky, I.O., Turnshek, 
             D.A., Nestor, D.B., \& Pettini, M. 2004, ApJ, 600, 613

\reference{} Scott, J., Bechtold, J., Dobrzycki, A., \& Kulkarni, V. 2000, 
             ApJS, 130, 67

\reference{} Scott, J., Bechtold, J., Morita, M., Dobrzycki, A., \& Kulkarni, 
             V. 2002, ApJ, 571, 665

\reference{} Sembach, K. R. \& Danks, A. C. 1994, A\&A, 289, 539


\reference{} Sofia, U.J. \& Meyer D.M. 2001, ApJ, 554, L221


\reference{} Steidel, C.C., Pettini, M., Dickinson, M., \& Persson, S.E.
             1994, AJ, 108, 2046

\reference{} Steidel, C.C., Adelberger, K. L., Giavalisco, M., Dickinson, M.,
             \& Pettini, M. 1999, ApJ, 519, 1

\reference{} Steidel, C.C., Kollmeier, J.A., Shapely, A.E., Churchill, C.W.,
             Dickinson, M., \& Pettini, M. 2002, ApJ, 570, 526

\reference{} Thurston, T. R., Edmunds, M. G., \& Henry, R. B. C. 1996, MNRAS, 
             283, 990

\reference{} Tremonti, C. A. \etal\ 2004, ApJ in press (astro-ph/0405537)

\reference{} Tully, R. B. \& Fisher, J. R. 1977, A\&A, 54, 661

\reference{} Turnshek, D.A., Rao, S., Nestor, D., Lane, W., Monier, E., 
	     Bergeron, J., \& Smette, A. 2001, ApJ, 553, 288

\reference{} Turnshek, D.A., Rao, S.M., Ptak, A.F., Griffiths, R.E., \& Monier,
             E.M. 2003, ApJ, 590, 730

\reference{} van der Hulst, Skillman, E., Smith, T., Bothun, G., McGaugh, S., 
             \& de Blok, W. 1993, AJ, 106, 548

\reference{} van Zee, L., Salzer, J. J., Haynes, M. P., O'Donoghue, A. A., \& 
             Balonek, T. J. 1998, AJ, 116, 2805

\reference{} Vladilo, G., Centuri\'on, M., Bonifacio, P., \& Howk, J. C. 2001, 
             ApJ, 557, 1007

\reference{} Vladilo, G. 2004, A\&A, 421, 479

\reference{} Vogt, N. P., Forbes, D. A., Phillips, A. C., Gronwall, C., Faber, 
             S. M., Illingworth, G. D., \& Koo, D. C. 1996, ApJ, 465, L15

\reference{} Welty, D. E., Morton, D. C., \& Hobbs, L. M. 1996, ApJS, 106, 533

\reference{} Wolfe, A.M., Briggs, F.H., \& Davis, M.M. 1982, ApJ, 259, 495

\reference{} Wolfe, A.M., Prochaska, J.X., \& Gawiser, E. 2003a, ApJ, 593, 215

\reference{} Wolfe, A.M., Gawiser, E., \& Prochaska, J.X. 2003b, ApJ, 593, 235

\reference{} Wright, E.L., \etal\ 1991, ApJ, 381, 200

\reference{} Zaritsky, D., Kennicutt, R.C. Jr., \& Huchra, J.P. 1994, ApJ,
             420, 87

\end{references}
\end{document}